\documentclass[floats,floatfix,superscriptaddress]{revtex4}

\usepackage{amsfonts,amsmath,amssymb}
\usepackage[pdftex]{graphicx}

\begin{document}

\title{Popularity versus similarity in growing networks}

\author{Fragkiskos Papadopoulos}

\affiliation{Department of Electrical Engineering, Computer Engineering and Informatics, Cyprus University of Technology, 33 Saripolou Street, 3036 Limassol, Cyprus}

\author{Maksim Kitsak}

\affiliation{Cooperative Association for Internet Data Analysis (CAIDA), University of California, San Diego (UCSD), La Jolla, CA 92093, USA}

\author{M.~\'Angeles Serrano}

\affiliation{Departament de F{\'\i}sica Fonamental, Universitat de Barcelona, Mart\'{\i} i Franqu\`es 1, 08028 Barcelona, Spain}

\author{Mari{\'a}n Bogu{\~n}{\'a}}

\affiliation{Departament de F{\'\i}sica Fonamental, Universitat de Barcelona, Mart\'{\i} i Franqu\`es 1, 08028 Barcelona, Spain}

\author{Dmitri Krioukov}

\affiliation{Cooperative Association for Internet Data Analysis (CAIDA), University of California, San Diego (UCSD), La Jolla, CA 92093, USA}

\begin{abstract}
{\em Popularity is attractive}~\cite{DoMe00pop}---this is the formula underlying preferential attachment~\cite{BarAlb99}, a popular explanation for the emergence of scaling in growing networks. If new connections are made preferentially to more popular nodes, then the resulting distribution of the number of connections that nodes have follows power laws~\cite{KraReLe00,DoMeSa00} observed in many real networks~\cite{Dorogovtsev10-book,Newman10-book}. Preferential attachment has been directly validated for some real networks, including the Internet~\cite{PaVaVe01,H.Jeongetal2003}. Preferential attachment can also be a consequence of different underlying processes based on node fitness, ranking, optimization, random walks, or duplication~\cite{DoMeSa01,BiBa01a,CaCaDeMu02,Vazquez2003,PaSaSo03,FoFl06,SoBo07,Motter2007}. Here we show that popularity is just one dimension of attractiveness. Another dimension is similarity. We develop a framework where new connections, instead of preferring popular nodes, optimize certain trade-offs between popularity and similarity. The framework admits a geometric interpretation, in which popularity preference emerges from local optimization. As opposed to preferential attachment, the optimization framework accurately describes large-scale evolution of technological (Internet), social (web of trust), and biological ({\it E.coli\/} metabolic) networks, predicting the probability of new links in them with a remarkable precision. The developed framework can thus be used for predicting new links in evolving networks, and provides a different perspective on preferential attachment as an emergent phenomenon.
\end{abstract}

\maketitle

More similar nodes have higher chances to get connected even if they are not popular. This effect is known as {\em homophily\/} in social sciences~\cite{McPh01,SiJe08}, and it has been observed in many real networks~\cite{Redner98,WatDoNew02,BoMaGo04-pnas,CraCo08,menczer02-pnas,menczer04-pnas}. In the Web~\cite{menczer02-pnas,menczer04-pnas}, for example, an individual creating her new homepage tends to link it not only to popular sites such as Google or Facebook, but also to not so popular sites that are close to her special interests, e.g., Tartini or free soloing. These observations suggest to introduce a measure of attractiveness which would somehow balance popularity and similarity.

The simplest proxy to popularity is the node birth time. All other things equal, older nodes have more chances to become popular and attract connections~\cite{KraReLe00,DoMeSa00}. If nodes join the network one by one, then the node birth time is simply the node number $t=1,2,\ldots$. To model similarity, we randomly place nodes on a circle abstracting the simplest similarity space. That is, the angular distances between nodes model their similarity distances, such as the cosine similarity or any other measure~\cite{menczer02-pnas,menczer04-pnas,CraCo08}. The simplest way to model a balance between popularity and similarity is then to establish new connections optimizing the product between popularity and similarity. In other words, the model is simply: (1)~initially the network is empty; (2) at time $t \geq 1$, new node $t$ appears at a random angular position $\theta_t$ on the circle; and (3)~connects to a subset of existing nodes $s$, $s<t$, consisting of the $m$ nodes with the $m$ smallest values of product $s\theta_{st}$, where $m$ is a parameter controlling the average node degree $\bar{k}=2m$, and $\theta_{st}$ is the angular distance between nodes $s$ and $t$ (Fig.~1(a,b)). At early times $t \leq m$, node $t$ connects to all the existing nodes.

\begin{figure*}
\includegraphics[width=5.5in]{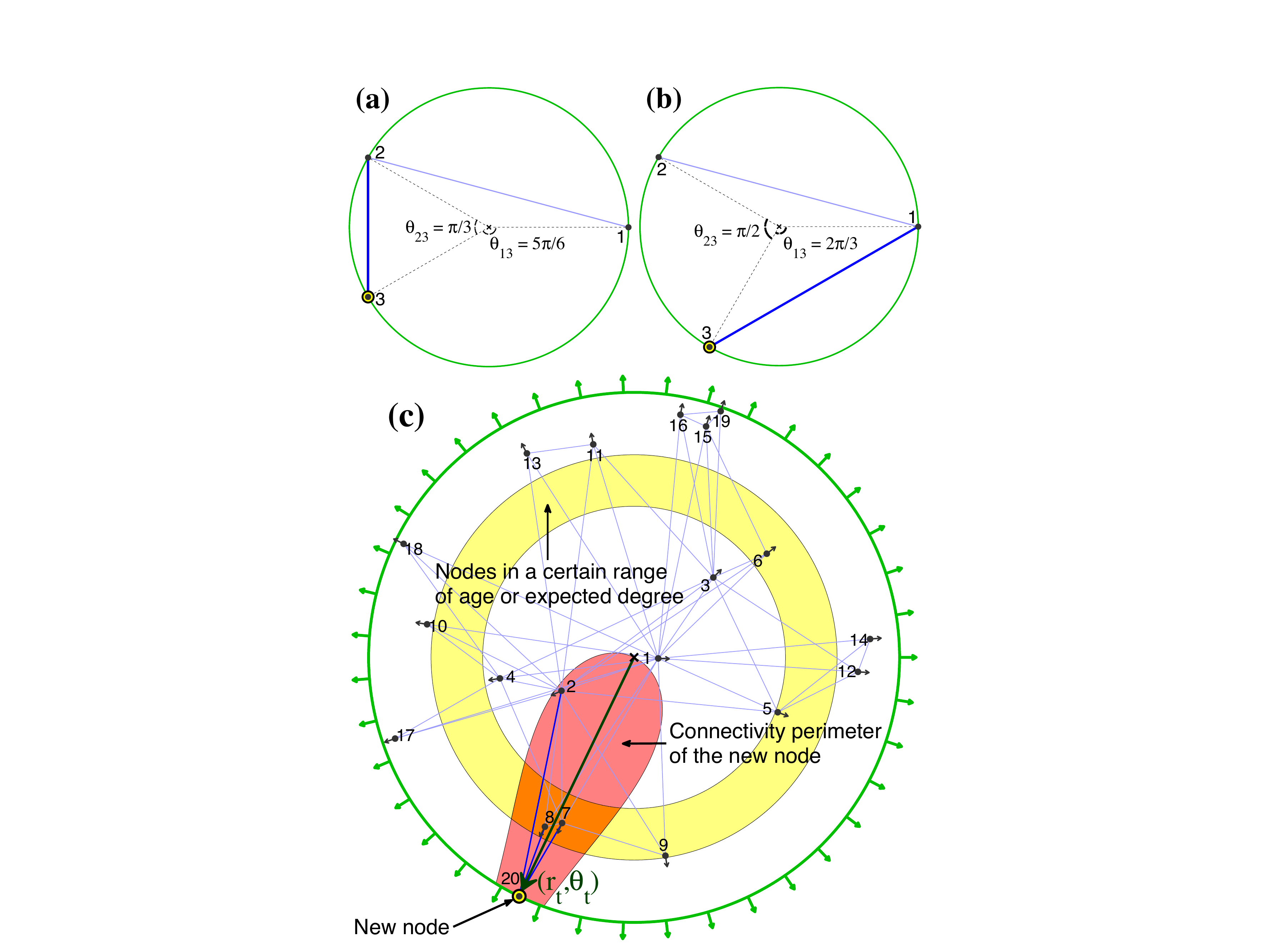}
\caption{Geometric interpretation of popularity$\times$similarity optimization. The nodes are numbered by their birth times, and located at random angular (similarity) coordinates. Upon its birth, the new circled node~$t$ connects to $m$ old nodes~$s$ minimizing $s\theta_{st}$.  The new connections are shown by the thicker blue links. In~(a,b) $t=3$ and $m=1$. In~(a) node~$3$ connects to node~$2$ because $2\theta_{23}=2\pi/3<1\theta_{13}=5\pi/6$. In~(b) node~$3$ connects to node~$1$ because $1\theta_{13}=2\pi/3<2\theta_{23}=\pi$. In~(c) an optimization-driven network with $m=3$ is simulated for up to $20$ nodes. The radial (popularity) coordinate of new node~$t=20$ is $r_t=\ln t$, and the node connects to the three hyperbolically closest nodes. The red shape marks the set of points located at hyperbolic distances less than $r_t$ from the new node. All nodes drift away from the crossed origin, emulating popularity fading as explained in the text. The drift speed in the shown network corresponds to the degree distribution exponent $\gamma=2.1$.}
\end{figure*}

This model finds an interesting geometric interpretation, shown in Fig.~1(c). Specifically, after mapping birth time
$t$ of a node to its radial coordinate $r_t$ via $r_t=\ln t$, all nodes lie not on a circle but on a plane---their
polar coordinates are $(r_t,\theta_t)$. It then turns out that new nodes connect simply to the {\em closest\/} $m$
nodes on the plane, except that distances are not Euclidean but hyperbolic~\cite{Bonahon09-book}. The hyperbolic distance between two
nodes at polar coordinates $(r_s,\theta_s)$ and $(r_t,\theta_t)$ is approximately
$x_{st}=r_s+r_t+\ln(\theta_{st}/2)=\ln(st\theta_{st}/2)$. Therefore the sets of nodes $s$ minimizing $x_{st}$ or $s\theta_{st}$ for each $t$ are identical.
The hyperbolic distance is then nothing but a convenient single-metric representation of a combination of the two attractiveness attributes, radial popularity and angular similarity. We will use this metric extensively below.

\begin{figure*}
\includegraphics[width=6.5in]{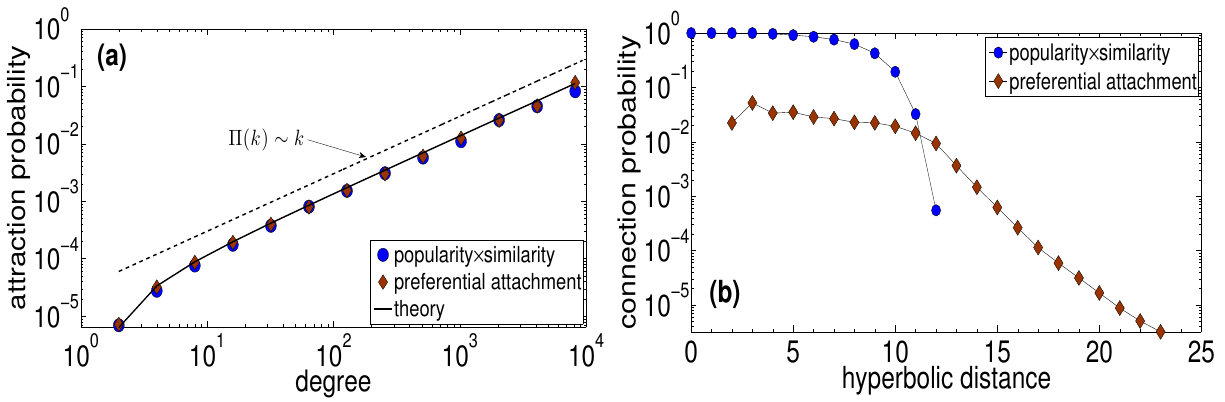}
\caption{Emergence of preferential attachment from popularity$\times$similarity optimization. Two
growing networks have been simulated up to $t=10^5$ nodes, one growing according to the
described optimization model, and the other according to PA. In both networks each new node connects to $m=2$ existing nodes.
The $\gamma\to2$ limit is not well-defined in PA, so that $\gamma=2.1$ is used instead as described in the text.
Plot~(a) shows the probability $\Pi(k)$ that an existing node of degree $k$ attracts a new link.
The solid line is the theoretical prediction. Plot~(b) shows the probability $p(x)$ that a pair of nodes located at
hyperbolic distance $x$ are connected. The average clustering (over all nodes) in the optimization and PA networks is $\bar{c}=0.83$
and $\bar{c}=0.12$, respectively.}
\end{figure*}

The networks grown as described may seem to have nothing in common with preferential attachment (PA)~\cite{BarAlb99,KraReLe00,DoMeSa00}.
Yet we show in Fig.~2(a) that the probability $\Pi(k)$ that an existing node of degree $k$ attracts a connection from a new node is the same linear function of $k$ in the described model and in PA. It is not surprising then that the degree distributions in PA and our model are the same power laws. In Section~\ref{sec:a1} we prove that the exponent $\gamma$ of this power law approaches $2$. Preferential attachment thus emerges as an effective process originating from optimization trade-offs between popularity and similarity.

However, there are crucial differences between such optimization and PA. In the latter, new nodes connect with the same probability $\Pi(k)$ to {\em any\/} nodes of degree $k$ in the network. In the former, new nodes connect only to specific {\em subsets\/} of such $k$-degree nodes that are closest to the new node along the similarity dimension $\theta$ (Fig.~1(c)). To quantify, we compare in Fig.~2(b) the probability of connection between a pair of nodes as a function of their hyperbolic distance in the two cases. We see that close nodes are almost always connected in the optimization model, while in PA the probability of their connections is lower by an order of magnitude. On the other hand, far apart nodes are never connected in the optimization model, as opposed to PA. These differences manifest themselves in the strength of clustering, which is the probability that two neighbors of the same node are connected. In PA, clustering is asymptotically zero~\cite{BoRi02}, while it is strong in many real networks~\cite{Dorogovtsev10-book,Newman10-book}. We show in Section~\ref{sec:a1} that the described optimization model leads to clustering that is strongest possible for networks with a given average degree and degree distribution.

Clustering and the power-law exponent can both be adjusted to arbitrary values via the following model modifications. We first consider the effect of popularity fading, observed in many real networks~\cite{AdHu00,Raan00}. We note that the closer the node to the center in Fig.~1(c), the more popular it is---the more new connections it attracts, and the higher its degree---providing the intuition behind the emergence of preferential attachment. Therefore to model popularity fading, we let all nodes drift away from the center such that the radial coordinate of node $s$ at time $t>s$ is increasing $r_s(t)=\beta r_s+(1-\beta)r_t$, where $r_s=\ln s$ and $r_t=\ln t$, and parameter $\beta\in[0,1]$. This modification is identical to minimizing $s^\beta\theta_{st}$ (or $s^b\theta_{st}^a$ with $\beta=b/a$) instead of $s\theta_{st}$. It changes the power-law exponent to $\gamma=1+1/\beta\geq2$. If $\beta=1$, the nodes do not move and $\gamma=2$. If $\beta=0$, all nodes move with the maximum speed, always lying on the circle of radius $r_t$, while the network degenerates to a random geometric graph growing on the circle. PA emerges at any $\gamma=1+1/\beta$ since the attraction probability $\Pi(k)$ is a linear function of degree $k$, $\Pi(k) \sim k+m(\gamma-2)$, the same as in PA~\cite{DoMeSa00}. We prove these statements in Sections~\ref{sec:a1}--\ref{sec:a4}, where we also show that the popular fitness model~\cite{BiBa01a} can be mapped to our geometric optimization framework by letting different nodes drift away with different speeds (Section~\ref{sec:a2}).

Since strongest clustering is due to connections to the closest nodes, to weaken clustering we allow connections to farther nodes. Connecting to the $m$ closest nodes is approximately the same as connecting to nodes lying within distance $R_t \sim r_t$, see Fig.~1(c) and Section~\ref{sec:a1}, where we derive the exact expression for $R_t$ fixing the average degree in the network. If new nodes $t$ establish connections to existing nodes $s$ with probability $p(x_{st})=1/[1+e^{(x_{st}-R_t)/T}]$, where parameter $T\geq0$ is the network temperature and $x_{st}$ is the hyperbolic distance between nodes $s$ and $t$, then clustering is a decreasing function of temperature. That is, temperature is the parameter controlling clustering in the network. At zero temperature, the connection probability $p(x_{st})$ is either $1$ or $0$ depending on whether distance $x_{st}$ is less or greater than $R_t$, so that we recover the strongest clustering case above, where new nodes connect only to the closest existing nodes. Clustering gradually decreases to zero at $T=1$, and remains asymptotically zero for any $T\geq1$~(Sections~\ref{sec:a1},~\ref{sec:zeta_extension}). At high temperatures $T\to\infty$ the model degenerates either to growing random graphs, or, remarkably, to standard PA~(Section~\ref{sec:a4}).

To investigate if similarity shapes the structure and dynamics of real networks as our model predicts, we consider a series of historical snapshots of the Internet, {\it E.coli\/} metabolic network, and the web of trust between people. The first two networks are disassortative, while the third is assortative, and its degree distribution deviates from power laws. We map these networks to their popularity$\times$similarity spaces (Methods Summary). The mapping infers the radial (popularity) and angular (similarity) coordinates for all nodes, so that we can compute the hyperbolic distances between all node pairs, and the probability of new connections as a function of the hyperbolic distance between corresponding nodes. This probability is shown in Fig.~3. It is close to the theoretical prediction by the model.

\begin{figure*}
\includegraphics[width=6.5in]{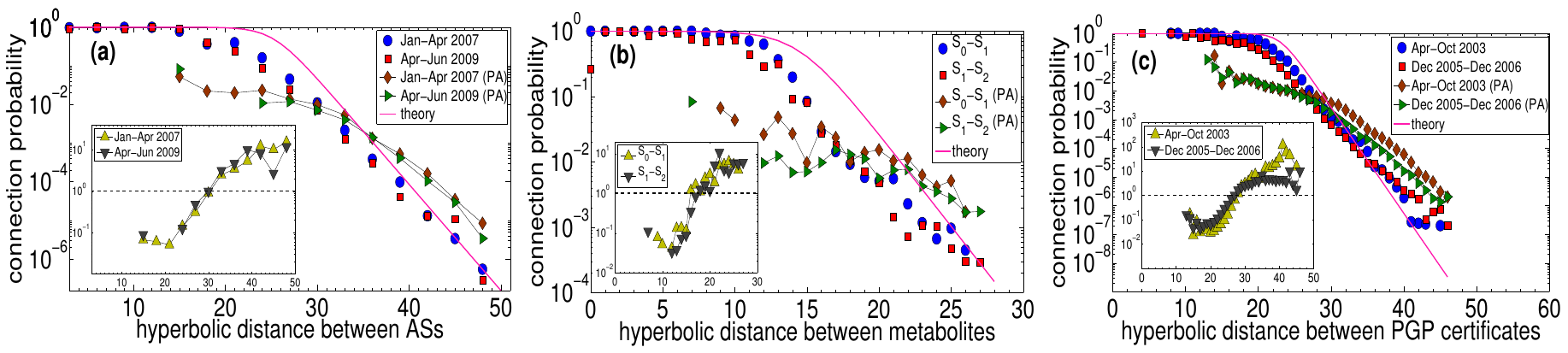}
\caption{Popularity$\times$similarity optimization in the growing Internet (plot~(a)), {\it E.coli\/} metabolic network (plot~(b)),
and Pretty-Good-Privacy (PGP) web of trust between people (plot~(c)). Each plot shows the probability of connections between new and old nodes, as a function of the hyperbolic (popularity$\times$similarity) distance~$x$ between them in the real networks (circles and squares) and in PA emulations (diamonds and triangles). To emulate PA, new links are disconnected from old nodes to which these links are connected in the real networks, and reconnected to old nodes according to PA. For a pair of historical network snapshots $S_0$ (older) and $S_1$ (newer), {\em new\/} nodes are the nodes present in $S_1$ but not in $S_0$, and {\em old\/} nodes are the nodes present both in $S_1$ and $S_0$. Each plot shows the data for two pairs of such historical snapshots. The solid curve in each plot is the theoretical connection probability in the optimization model with the parameters corresponding to a given real network. Since the probability of new connections in the real networks is close to the theoretical curves, the shown data demonstrate that these networks grow as the popularity$\times$similarity optimization model predicts, while PA, accounting only for popularity, is off by orders of magnitude in predicting the connections between similar (small $x$) or dissimilar (large $x$) nodes. To quantify this inaccuracy, the insets show the ratio between the connection probabilities in PA emulations and in the real networks, i.e., the ratios of the values shown by diamonds and circles, and by triangles and squares in the main plots. The $x$-axes in the insets are the same as in the main plots.}
\end{figure*}

This finding is important for several reasons. First, it shows that real-world networks evolve as our framework predicts. Specifically, given the popularity and similarity coordinates of two nodes, they link with probability close to the theoretical in the model. The framework may thus be used for link prediction, a notoriously difficult and important problem in many disciplines~\cite{ClMo08}, with applications ranging from predicting protein interactions or terrorist connections to designing recommender and collaborative filtering systems~\cite{MeEl11}. Second, Fig.~3 {\em directly\/} validates our framework and its core mechanism. It is not surprising then that, as a consequence, the synthetic graphs that the model generates are remarkably similar to real networks across a range of metrics~(Section~\ref{sec:dk_compare}), implying that the framework can be also used for veracious modeling of real network topologies. We review related work in Section~\ref{sec:related_work}, and to the best of our knowledge, there is no model that would simultaneously: (1)~be simple and universal, i.e., applicable to many different networks, (2)~have a similarity space as its core component, (3)~cast PA as an emergent phenomenon, (4)~generate graphs similar to real networks across a wide range of metrics, and (5)~validate the proposed growth mechanism directly. Validation is usually limited to comparing certain graph metrics, such as degree distribution, between modeled and real networks, which ``validates'' a consequence of the mechanism, not the mechanism itself. Direct validation is usually difficult because proposed mechanisms tend to incorporate many unmeasurable factors---economic or political factors in Internet evolution, for example. We cannot measure all the factors or node attributes contributing to node similarity in any of the considered real networks either. Yet, the angular distances between nodes in our approach can be considered as projections of properly weighted combinations of all such similarity factors affecting network evolution, and we can infer these distances using statistical inference methods, directly validating the growth mechanism.

To summarize, popularity is attractive, but so is similarity. Neglecting the latter would lead to severe aberrations. In the Internet, for example, a local network in Nebraska would connect directly to a local network in Tibet, the same way as in the Web, a person not even knowing Tartini or free soloing would suddenly link her page to these subjects. The probability of such dissimilar connections is very low in reality, and the stronger the similarity forces, the smaller this probability is. Neglecting the network similarity structure leads to overestimations or underestimations of the probability of dissimilar or similar connections by orders of magnitude (Fig.~3). However, one cannot tell the difference with preferential attachment by examining node degrees only. The probability that an existing node of degree~$k$ attracts a new link optimizing popularity$\times$similarity is exactly the same linear function of $k$ as in preferential attachment (Fig.~2(a)). Figure~S1 shows that this function is indeed realized in the considered real networks, re-validating {\em effective\/} preferential attachment for these networks. Therefore the popularity$\times$similarity optimization approach provides a natural geometric explanation for the following ``dilemmas'' with preferential attachment. On one hand, preferential attachment has been validated for many real networks, while on the other hand, it requires exogenous mechanisms to explain not only strong clustering, but also linear popularity preference, and how such preference can emerge in real networks, where nodes do not have any global information about the network structure. Since preferential attachment appears as an {\em emergent phenomenon\/} in the framework developed here, this framework provides a simple and natural resolution to these dilemmas, and this resolution is directly validated against large-scale evolution of drastically different real networks. We conclude with the observation that the knowledge of exactly the closest nodes in the hyperbolic popularity$\times$similarity space does require the precise global information about all node locations. However, non-zero temperatures smooth out the sharp connectivity perimeter threshold in Fig.~1(c), thus modeling reality where this proximity information is not precise and mixed with errors and noise. In that respect, preferential attachment is a limiting regime with similarity forces reduced to nothing but noise.

\section*{Methods Summary}

To infer the radial $r_i$ and angular $\theta_i$ coordinates for each node $i$ in a real network snapshot with adjacency matrix $a_{ij}$, $i,j=1,2,\ldots,t$, we use the Markov Chain Monte Carlo (MCMC) method described in detail in Section~\ref{sec:mapping}. Specifically, we derive there the exact relation between the expected current degree $\overline{k_i}$ of node $i$ and its current radial coordinate $r_i$, which scales as $\overline{k_i}\sim e^{r_t-r_i}$. To infer the radial coordinates we use the same expression substituting in it the real degrees $k_i$ of nodes instead of their expected degrees. Having the radial coordinates inferred, we then execute the Metropolis-Hastings algorithm to find the node angular coordinates that maximize likelihood ${\cal L}=\prod_{i<j}p(x_{ij})^{a_{ij}}[1-p(x_{ij})]^{1-a_{ij}}$, where $p(x_{ij})=1/[1+e^{(x_{ij}-R)/T}]$ is the connection probability in the model, and parameters $R$ and $T$ are defined by the average node degree and clustering in the network via expressions in Section~\ref{sec:a1}. Likelihood ${\cal L}$ is the probability that the network snapshot with node coordinates $(r_i,\theta_i)$, defining the hyperbolic distances $x_{ij}$ between all nodes, is produced by the model. The algorithm employs an MCMC process which finds coordinates $\theta_i$ for all $i$ that approximately maximize ${\cal L}$. Further details are in Sections~\ref{sec:mapping},~\ref{sec:fitting}, where we also show that the method yields meaningful results for the considered networks, but not for a network (movie actor collaborations) to which popularity$\times$similarity optimization does not apply.

In Fig.~3, the nodes in plots~(a), (b), and (c) are Autonomous Systems (ASs), metabolites, and PGP certificates of people. Parameters $(R,T)$ used to infer the coordinates and to draw the theoretical connection probability are $(25.2,0.79)$, $(14.4, 0.77)$, and $(23, 0.59)$. Each plot shows data for two pairs of snapshots: plot~(a) January, April, 2007, and April, June, 2009; plot~(b) $S_0$, $S_1$, and $S_1$, $S_2$ defined in Section~\ref{sec:real_nets}; and plot~(c) April, October, 2003, and December $2005$, December $2006$. Few missing data points in the empirical curves (circles and squares) indicate that there are no node pairs at the corresponding distances after the mapping, whereas {\em extra\/} missing points in the PA emulation curves (diamonds and triangles) indicate that all node pairs at those distances are not connected after PA emulations, meaning that the PA connection probability is zero there.

\setcounter{figure}{0}
\makeatletter
\renewcommand{\thefigure}{S\@arabic\c@figure}

\section{Real-world networks}
\label{sec:real_nets}

Here we provide details on the real-world network data used to validate the popularity$\times$similarity optimization approach. We have considered the AS Internet, the {\it E.coli\/} metabolic network, and the web of trust among people extracted from Pretty-Good-Privacy (PGP) data. That is, we have validated our approach against three paradigmatic real networks, from three different domains---technology, biology, and society.

\begin{figure*}
\includegraphics[width=7in, height=1.6in]{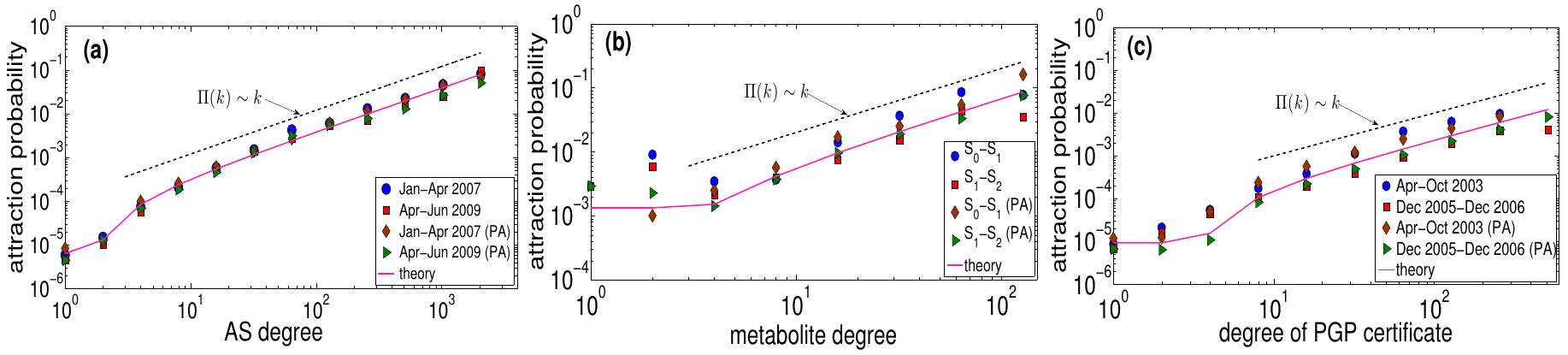}
\caption{Plots~(a),~(b),~(c) show the probability $\Pi(k)$ that an old node attracts a new link in
the AS Internet, the {\it E.coli\/} metabolic network, and the PGP web of trust, respectively. The network snapshots in each case
are the ones used in Fig.~3. The plots also show the results for the corresponding PA emulations and the theoretical prediction, which is the same linear function in popularity$\times$similarity optimization and PA.}
\end{figure*}

\subsection{Internet}
\label{sec:internet_data}

The Internet data used in Fig.~3(a) and Fig.~S11 of Section~\ref{sec:dk_compare} is collected and prepared as follows. First, we obtain $11$ lists of all the autonomous systems (ASs) observed in a collection of Border Gateway Protocol (BGP) data exactly as described in~\cite{DhDo08}. These AS lists are linearly spaced in time with the interval of three months: time $t=0$ corresponds to January 2007, $t=1$ is April 2007, and so on until $t=10$, June 2009. We denote the obtained AS lists by $L_t$. For any pair of $t$ and $t'>t$, we call the ASs present both in $L_t$ and $L_{t'}$ the {\em old\/} ASs, and the ASs present only in $L_{t'}$ but not in $L_t$ are called the {\em new\/} ASs. The number of ASs in $L_0$ is $17258$, while the numbers of new ASs in $L_{t'}$ with $t'=1,2,\ldots,10$ compared to $t=0$ are $806$, $1614$, $2389$, $3103$, $3973$, $4794$, $5434$, $5843$, $6207$, and $6426$. We then take the Archipelago AS topology~\cite{ClHy09} of June 2009, available at \cite{as_topo_data}, and for each $t=0,1,\ldots,10$ we remove from it all ASs and their adjacent links that are not in $L_t$, thus obtaining a time series of historical AS topology snapshots $S_t$. We then map each $S_t$ to the hyperbolic space as described in Section~\ref{sec:mapping}, and for each $t=0,1,\ldots,9$ and $t'=t+1$ we compute the empirical probability~$p(x)$ of connections between new and old ASs as a function of hyperbolic distance $x$ between the ASs. To compute~$p(x)$, we linearly bin distance~$x$, and show in each bin the ratio of the number of connected ASs to the total number of AS pairs located at hyperbolic distances falling within this bin. To avoid clutter, Fig.~3(a) shows the results for the first and last pairs of consecutive snapshots, i.e., $S_0, S_1$ and $S_9, S_{10}$. The number of new ASs in each pair is respectively $806$ and $259$. Similar results hold for all intermediate snapshot pairs. The figure also shows the results of PA emulations. To emulate PA in a snapshot pair, the links adjacent to a new AS are first disconnected from the old ASs to which these links are connected in reality, and then reconnected to old ASs chosen randomly with the normalized probability $\sim k+\bar{k}(\gamma-2)/2$, where $k$ is the number of connections the AS has to other old ASs, and $\bar{k}=5.3$, $\gamma=2.1$, taken from the Internet. The average clustering in the Internet is $\bar{c}=0.61$, and both $\bar{k}$ and $\bar{c}$ are stable across the considered period. Figure~S1(a) validates effective preferential attachment for the pairs of Internet snapshots considered in Fig.~3(a).

\subsection{{\it E.coli\/} metabolic network}
\label{sec:metabolic_data}

We use the bipartite metabolic network representation of the {\it E.coli\/} metabolism from~\cite{Serrano:2011}, reconstructed from data in the BiGG database~\cite{Schellenberger:2010,Bigg}, {\it i}AF1260 version of the K12 MG1655~\cite{Feist:2007} strain. The bipartite representation differentiates two subsets of nodes, metabolites and reactions, mutually interconnected through unweighted and undirected links, without self-loops or dead end reactions. Reactions that do not involve direct chemical transformations, such as diffusion and exchange reactions, are avoided and isomer metabolites are differentiated. To enhance the resolution of the mapping procedure, currency metabolites are eliminated (h, h2o, atp, pi, adp, ppi, nad, nadh, amo, nadp, nadph), altogether with a few isolated reaction-metabolite pairs and reaction-metabolite-reaction triplets. This leads to a globally connected set of $1512$ reactions and $1010$ metabolites. Starting from this bipartite network, we construct its one mode projection over the space of metabolites, that is, we consider only metabolites and declare two metabolites as connected if they participate in the same reaction in the original bipartite network. The resulting unipartite network of metabolites has a power law degree distribution with exponent $\gamma=2.5$, average degree $\bar{k}=6.5$, and the average clustering is $\bar{c}=0.48$.

Empirical data for ancestral metabolic networks is not available. However, it has been argued that there exists a direct relation between the evolutionary history of metabolism and the connectivity of metabolites. The hypothesis is that metabolic networks grew by adding new metabolites, such that the most highly connected metabolites should also be the phylogenetically oldest \cite{Morowitz:1999,FellWagner:2000,FellWagner:2000b,Wagner:2002}. Following this idea, we sorted the network of metabolites by degree to construct an ancestor core metabolic network of $460$ metabolites with degrees larger than $4$, and two shells including metabolites of degrees $4$ and $3$, respectively. Each shell is meant to represent the addition of new metabolites in subsequent evolutionary steps. The first shell consists of $142$ new metabolites and the second shell of $171$ new metabolites. Time $t=0$ corresponds to the core network $S_0$. Time $t=1$ corresponds to the snapshot of the topology $S_1$ consisting of the metabolites in $S_0$ and the new metabolites in the first shell. And, time $t=2$ corresponds to the snapshot of the topology $S_2$ consisting of the metabolites in $S_1$ and the new metabolites in the second shell. We map $S_0, S_1, S_2$ to the hyperbolic space and compute the empirical connection probability, following the same procedure as in the previous subsection for the Internet. As before, we also perform PA emulations. The results are shown in Fig.~3(b) and in Fig.~S1(b), and are very similar to those of Figs.~3(a),~S1(a). The data from this section are also used in Fig.~S12 of Section~\ref{sec:dk_compare}.

\subsection{PGP web of trust}
\label{sec:pgp_data}

Pretty-Good-Privacy (PGP) is a data encryption and decryption computer program that provides cryptographic privacy and
authentication for data communication~\cite{openpgp}. PGP web of trust is a directed network where nodes are
certificates consisting of public PGP keys and owner information. A directed link in the web of trust pointing from
certificate {\it A} to certificate {\it B} represents a digital signature by owner of {\it A} endorsing the
owner/public key association of {\it B}. We use temporal PGP web of trust data collected and maintained by J\"{o}rgen
Cederl\"{o}f~\cite{cederlof}.

The PGP web of trust (WoT) data is analyzed as follows. We consider two closely spaced in time pairs of WoT
snapshots taken in April 2003, October 2003, and December 2005, December 2006. For each of the directed graphs we
form their undirected counterparts by taking into account only bi-directional trust links  between the certificates.
For each of the undirected counterparts we isolate its largest connected component. Then, for each pair of the snapshots we
identify old and new sets of nodes. As before, the set of old nodes contains all nodes present in both snapshots, while
the set of new nodes contains nodes that are present in the newer snapshot and not in the older snapshot. We refer to the obtained
undirected connected subgraphs of the WoT, as snapshots $S_0, S_1, S_2, S_3$,
for April 2003, October 2003, December 2005, December 2006, respectively. The numbers of nodes in $S_0, S_1, S_2, S_3$ are
$14367, 17155, 23797, 26701$, while the average degree $\bar{k}$ is $5.3, 6.2, 7.9, 8.1$ and the average clustering is $\bar{c}=0.47$-$0.48$.
The degree distribution can be roughly approximated by a power-law with exponent $\gamma=2.1$, yet we observe some deviations from this
power at high degrees, see Fig.~S13(a). We map $S_0, S_1, S_2, S_3$ to the hyperbolic space
and compute the empirical connection probability, following the same procedure as in the previous two subsections. As before, we
again perform PA emulations. The results are shown in Fig.~3(c) and in Fig.~S1(c), and are very similar to Figs.~3(a),~3(b) and
Figs.~S1(a),~S1(b). The data from this section are also used in Fig.~S13 of Section~\ref{sec:dk_compare}.

By using the PGP data as described, we strengthen the social component of the WoT, since we only consider bi-directional
signatures, i.e., pairs of users (owners of PGP keys) who have reciprocally signed each other's keys. This filtering process increases the probability that the
connected users know each other, and makes the extracted network a reliable proxy to the underlying social network. We consider the PGP WoT since it is a massive
evolving unipartite graph, which represents real social relationships of trust among individuals, and for which complete historical data is available.

\section{Inferring the popularity and similarity coordinates}
\label{sec:mapping}

Here we describe the network mapping method used to infer the popularity and similarity coordinates in the considered real networks.

Given a snapshot of a real network consisting of $t$ nodes, we use the Markov Chain Monte Carlo (MCMC) method described in~\cite{BoPa10} to compute the current radial (popularity) $r_s(t)$ and angular (similarity) $\theta_s$ coordinates for each node $s$ in the network. In this section we briefly describe this method. See~\cite{BoPa10} for further details.

To infer the radial coordinates is relatively easy. In Section~\ref{sec:a1}, we derive the exact relation between the expected current degree $\overline{k_s(t)}$ of node $s$ and its current radial coordinate $r_s(t)$ in the model, $\overline{k_s(t)}\sim e^{r_t-r_s(t)}$, where $r_t$ is the current radius of the hyperbolic disc. Therefore to infer the radial coordinates in a real network, we use the same expression substituting in it the real degrees $k_s(t)$ of nodes instead of their expected degrees.

The inference of the angular coordinates is much more involved. In summary, we first measure the average degree, power-law exponent, and average clustering in the network to determine $m$, $\beta$, and $T$, and then execute the Metropolis-Hastings algorithm trying to find the angular coordinates that would maximize the probability (or likelihood)
\begin{eqnarray}
\label{eq:L}
{\cal L}&=&\prod_{i<j}p(x_{ij})^{a_{ij}}[1-p(x_{ij})]^{1-a_{ij}},\\
p(x_{ij})&=&\frac{1}{1+e^{\frac{x_{ij}-r_t}{T}}},
\end{eqnarray}
that a given real network with adjacency matrix $a_{ij}$, $i,j=1,2,\ldots,t$, and with given node coordinates defining the hyperbolic distances $x_{ij}$ between nodes, is produced by the model with the measured parameters. The algorithm operates by repeating the following steps:
\begin{enumerate}
\item  Compute the current likelihood ${\cal L}_c$;
\item  Select a random node;
\item  Move it to a new random angular location;
\item  Compute the new likelihood ${\cal L}_n$;
\item  If ${\cal L}_n > {\cal L}_c$, accept the move;
\item  Otherwise, accept the move with probability ${\cal L}_n/{\cal L}_c$;
\end{enumerate}
and some manual intervention and guidance are needed for this algorithm to actually succeed in a reasonable amount of computing time~\cite{BoPa10}.

Given a historical series of real network topology snapshots $S_0, S_1, S_2,...$, we first map $S_0$ to the hyperbolic space exactly as just described, i.e., we compute the current radial coordinate, and angular position of each node. In the previous section, we have considered a series of $11$ AS Internet snapshots $S_0, S_1,...,S_{10}$, a series of $3$ {\it E.coli\/} metabolic network snapshots $S_0, S_1, S_2$ and two series of two PGP web of trust snapshots $S_0, S_1$ and $S_2, S_3$. For new nodes in consecutive snapshots of a series we compute their hyperbolic coordinates keeping the coordinates of old nodes fixed. That is, once a node appears at some time and gets its coordinates computed, its coordinates never change. Although according to the model the radial coordinates of nodes should increase with time (unless $\gamma=2$), here, for simplicity, we keep them fixed. This simplification is justified because the difference $\Delta r=\left(\frac{\gamma-2}{\gamma-1}\right)\ln{\frac{t+\Delta t}{t}}$ in the radial coordinate of every node in snapshots with $t+\Delta t$ and $t$ nodes is not significant in the closely spaced snapshot series that we consider. In particular, the maximum value of $\Delta r$ in the Internet, metabolic and PGP snapshot series is respectively $0.058, 0.35, 0.032$. Another simplification is that since the old node coordinates are fixed, we compute the angular coordinate for new nodes $i$ using only their local contributions to the total likelihood in Equation~(\ref{eq:L}), i.e., instead of~(\ref{eq:L}) we use
${\cal L}_i=\prod_{j \neq i}p(x_{ij})^{a_{ij}}[1-p(x_{ij})]^{1-a_{ij}}$.

Figure~3 shows that the empirical connection probabilities between new and old nodes in the Internet, {\it E.coli\/} metabolic network, and PGP web of trust, follow their theoretical predictions. These results signify that new connections in these networks are established as our popularity$\times$similarity optimization framework predicts.

\section{Discussion of the mapping method}
\label{sec:fitting}

Here we show that the mapping method yields meaningful results, without overfitting or other artifacts.

The number of parameters in the model is large. It is proportional to the network size, since we have to infer coordinates for each node. Therefore a natural question that arises is whether the mapping method described above yields meaningful results. In particular, could it be the case that the good match between empirical and theoretical connection probabilities in Fig.~3 is due to overfitting?

In this section we show that the inference results are indeed meaningful, since we find strong correlations between inferred coordinates and network-specific node attributes in each considered network. We also compute the logarithmic loss, which is the metric of the inference quality for statistical inference methods based on maximum-likelihood estimation. We show that this quality is good for each considered network, confirming that the results in Fig.~3 are not an (overfitting) artifact. Finally, we provide an example of real network (IMDb), where this quality is poor, and so is the logarithmic loss. Collectively, these results show that the inference method does not suffer from overfitting. In particular, if it were the case, then this method would yield statistically good results for {\em any} network.

\subsection{The mapping yields meaningful results}

\subsubsection{Internet}

In~\cite{BoPa10}, where we study Internet routing, we use the method described in Section~\ref{sec:mapping} to map the Archipelago AS topology of June 2009, used in Section~\ref{sec:internet_data}. The mapping yields meaningful results, since ASs belonging to the same country are mapped close to each other, see Figs.~3 and 5 in~\cite{BoPa10}. More precisely, one can see from Figs.~3,5 in~\cite{BoPa10} that for the majority of countries, their ASs are localized in narrow angular regions.
That is, even though the mapping method is completely geography-agnostic, it discovers meaningful groups or communities of ASs belonging to the same country.

The reason for this effect is that ASs belonging to the same country are usually connected more densely to each other than to the rest of the world, and the method correctly places all such ASs in narrow regions close to each other. We can also see from Fig.~3 in~\cite{BoPa10} that in many cases, geographically or politically close countries are located close to each other on the circle. These results prove that the angular coordinates inferred by the method reflect reality well, as is the case with the other two networks that we consider here.

\subsubsection{{\it E.coli\/} metabolic network}

Distances in metabolic network maps give a measure of the chemical potential of metabolites to participate jointly in reactions, such that higher reaction likelihoods are naturally associated to metabolites which are closer in the underlying space. It is then expected that metabolites participating in reactions in the same biochemical pathway would cluster in specific regions in the inferred space.

This was indeed observed in~\cite{Serrano:2011} for the cartographic network representation of the metabolism of {\it E.coli\/}, see Figs.~2, 3 and 4 there.
The geometric embedding of the metabolic network in Fig.~2 in~\cite{Serrano:2011}, obtained by the same mapping method we use in this paper, shows that metabolites that participate jointly in reactions are mapped close to each other, i.e., in the same angular regions. In particular, pathways---classically understood as chains of step-by-step reactions which transform a principal chemical into another---are in general strongly localized, even though some adopt either a discrete bi-modal or a multi-peaked form, and only a very small fraction transversally spread over the circle (Fig.~3 in~\cite{Serrano:2011}).
Furthermore, pathways in related functional categories tend to concentrate into well defined sectors (Fig.~4 in~\cite{Serrano:2011}). Therefore our model discriminates well the concentrated pathways, most frequent and consistent with the classical view of modular subsystems, from others, formed of subunits, and even from those responsible of producing or consuming metabolites used extensively in many other pathways.

\subsubsection{PGP web of trust}

To see if the mapping method yields meaningful results in the case of the PGP web of trust (WoT), we consider its mapped topology of April 2003 from Section~\ref{sec:pgp_data}. For each node (PGP certificate) in the topology, the data we use~\cite{cederlof} contain the
email address of the corresponding owner of the PGP certificate. For each email address we can identify the top-level domain that the email address belongs to, which is
the last part of the email address. Examples of top-level domains are .com, .net, .org, .de, .fr., .it and other country codes. Therefore, for each node
in the PGP network we can identify the top-level domain that the node belongs to. If the inferred angular coordinates reflect reality well, then we expect
PGP nodes belonging to the same country code to be mapped to angular locations close to each other, since in general people in the WoT are expected to trust other people from their own country more. By contrast, we do not expect this to be the case for generic top-level domains such as .net. In Fig.~S15
we show that the mapping method indeed yields meaningful results, as expected.

\setcounter{figure}{14}
\begin{figure*}
\centerline{\includegraphics [width=6.0in]{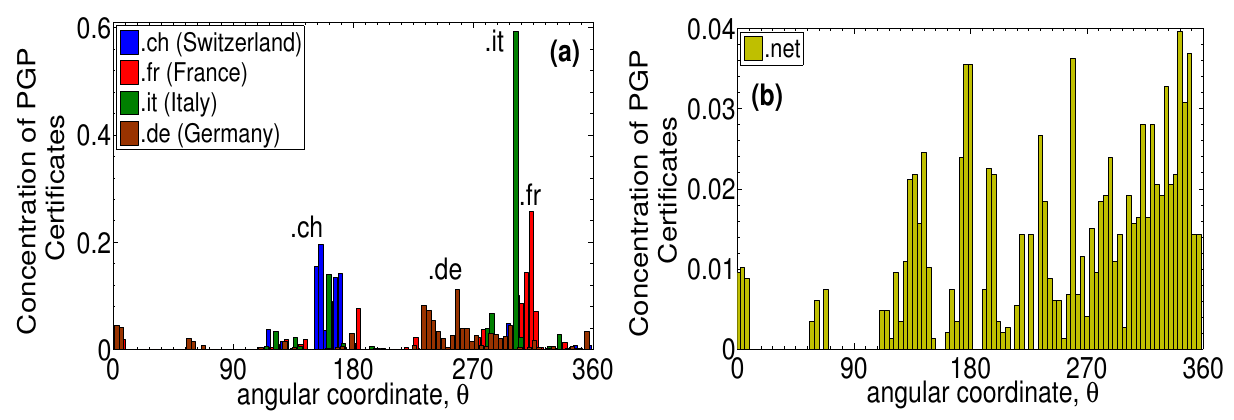}}
\caption{The mapping of the PGP web of trust yields meaningful results since PGP certificates belonging to the same
country code are mapped close to each other (plot~(a)). Similar results hold for other country codes. By contrast, certificates belonging to the generic top-level domain .net are widespread (plot~(b)), as expected. In~\cite{BoPa10} we show that the method also yields meaningful results for the Internet, and in~\cite{Serrano:2011} for the {\it E.coli\/} metabolic network.}
\end{figure*}

\subsection{Overfitting considerations and logarithmic loss}

\subsubsection{The number of parameters versus the number of predictions}

In general, a statistical inference method may suffer from overfitting if the number of parameters in the model is comparable or larger than the number of predicted parameters. Here we show that for any reasonably sized network, the former is much smaller than the latter in our model.

Indeed,
given a snapshot of a real network consisting of $t$ nodes, the mapping method in Section~\ref{sec:mapping} finds the angular coordinate of every node in the network such that the likelihood that the network is produced by the model is maximized. That is, if there are $t$ nodes in the network the method infers $t$ parameters (angular coordinates).

However, we stress that if a network consists of $t$ nodes, then there are $O(t^2)$ node pairs in the network, and for each node pair $i, j \leq t$, the model predicts an independent probability of the existence of a link between this node pair $p(x_{ij})$. If the mapping is successful, then \emph{every} node pair $i, j \leq t$ is placed at the right hyperbolic distance. In other words, if the fitting is successful, then with \emph{just} $t$ parameters, the model manages to successfully make $O(t^2)$ predictions.

If we consider new nodes in a subsequent snapshot, the method infers their angular coordinates such that they are all placed at the right hyperbolic distances with respect to the old nodes. If there are $\Delta t$ new nodes and $t$ existing nodes, and the mapping is successful, then with \emph{just} $\Delta t$ parameters, the model makes $O(t\Delta t)$ predictions.

If a real network is well described by the model, then the fitting of the large number of unknowns with a significantly smaller number of parameters is expected to be successful, as Fig.~3 illustrates. However, to compute the empirical connection probability in Fig.~3, we have to bin the hyperbolic distances into a small number of bins to have statistically reliable results for ratios of the number of connected node pairs to the total number of node pairs at distances within each bin. Instead of a large ensemble of graphs generated with the same parameters, we have only one real network, and we do not have any other method to compute the empirical connection probability for it. Therefore it is desirable to assess the mapping quality using an appropriate metric independent of any binning. Such metric for maximum-likelihood inference methods is logarithmic loss.

\subsubsection{Logarithmic loss}
\label{sec:further_tests}

In general, the logarithmic loss~\cite{Cesa-Bianchi2006} is defined as
\begin{equation}
L \equiv - \log\mathcal{L},
\end{equation}
where $\mathcal{L}$ is likelihood. Since maximum-likelihood inference methods operate by maximizing likelihood, logarithmic loss is a natural metric of the quality of the results that these methods produce. Specifically, if the results are good, then logarithmic loss is small. To estimate how small is ``small'' here, one usually compares against the case with random parameter assignments.

In our case, the likelihood ${\cal L}$ is defined in Equation~(\ref{eq:L}). That is, for a given real network and a given set of inferred coordinates, the logarithmic loss is
\begin{equation}
L \equiv - \sum_{i \neq j} \left[a_{ij} \log \left[p(x_{ij})\right ] + (1- a_{ij}) \log \left[1-p(x_{ij})\right]\right],
\end{equation}
where the above sum goes over all $O(t^2)$ pairs of nodes $i, j$, where $t$ is the network size. That is, we stress that logarithmic loss depends on {\em all} the $O(t^2)$ predicted probabilities $p(x_{ij})$. The logarithmic loss is nothing but the absolute value of the logarithm of the probability that the network is generated by the model, given the set of inferred node coordinates.

We compute logarithmic losses
for the Internet, {\it E.coli} metabolic network, and the PGP web of trust, with the node coordinates inferred by our mapping method. We contrast these
logarithmic losses against those obtained for the same networks with random angular coordinates. That is, we first assign to each node an angular coordinate
drawn uniformly at random from $[0,2\pi]$. The randomized logarithmic loss is then
\begin{equation}
L_{rand} \equiv - \sum_{i \neq j} \left[a_{ij} \log \left[p(\tilde{x}_{ij})\right ] + (1- a_{ij}) \log \left[1-p(\tilde{x}_{ij})\right]\right],
\end{equation}
where $\tilde{x}_{ij}$ is the hyperbolic distance between nodes $i$ and $j$ with random angular coordinates.
The smaller the $L$ compared to $L_{rand}$, the better the quality of the mapping, i.e., the better our model
describes a given real network.

To test the robustness of the inferred coordinates we also
calculate logarithmic losses after distorting inferred angular coordinate $\theta_i$ to
\begin{equation}
\tilde{\theta}_{i} = \theta_i + \delta \epsilon,
\end{equation}
where $\delta = 0.05, 0.1$ radians and $\epsilon$ a random variable drawn uniformly from the interval $[-1,1]$.

The logarithmic loss values are reported in Table~\ref{tab:si_table1}. From the table we observe that the logarithmic losses
calculated using the inferred angular coordinates are significantly smaller than those with random angular coordinates, indicating that
the considered real networks are well described by our model, corroborating the results in Fig.~3.

\begin{table}
\footnotesize
\begin{center}
\begin{tabular}{|c|c|c|c|c|c|c|c|}
\hline Network Name & $L$ & $\tilde{L}$, $\delta=0.05$ & $\tilde{L}$, $\delta=0.1$ & $L_{rand}$ & ${{\mathcal L}/{\mathcal L_{rand}}}$\\
\hline Internet (April 2007) & $1.4\times10^{5}$ & $1.8\times10^{5}$ & $2.3\times10^{5}$ & $2.7\times10^{5}$ & $\exp(1.3\times10^{5})$ \\
\hline {\it E.coli} metabolic ($S_0$) & $7.3\times10^{3}$ & $9.0\times10^{3}$ & $9.6\times10^{3}$ & $1.4\times10^{4}$ & $\exp(6.7\times10^{3})$\\
\hline PGP web of trust (April 2003) & $6.9\times10^{4}$ & $1.7\times10^{5}$ & $2.4\times10^{5}$ & $3.0\times10^{5}$ & $\exp(2.3\times10^{5})$\\
\hline
\end{tabular}
\caption{Logarithmic losses calculated with the inferred coordinates, distorted angular coordinates with $\delta = 0.05, 0.1$ radians, and fully
randomized angular coordinates, for the Internet (April 2007 snapshot), {\it
E.coli} metabolic network ($S_0$ snapshot), and PGP web of trust (April 2003 snapshot). The last column shows the ratios of likelihoods
$\mathcal{L} / \mathcal{L}_{rand} = \exp(L_{rand} - L)$, which are the ratios of the probability that the network is produced by the model with
the inferred angular coordinates, to the same probability with all these coordinates being random.
\label{tab:si_table1}}
\end{center}
\end{table}

We also compute logarithmic losses
considering only the links between new and old nodes. That is, given two consecutive snapshots of a network $S_{t}$ and $S_{t+1}$, we define the logarithmic loss as
\begin{equation}
L \equiv - \sum_{i , j} \left[a_{ij} \log \left[p(x_{ij})\right ] + (1- a_{ij}) \log \left[1-p(x_{ij})\right]\right],
\end{equation}
where summation is now over only $O(t\Delta t)$ new-old node pairs, and where $t$ is the number of old nodes, and $\Delta t$ the number of new nodes. Again,
the logarithmic losses using the inferred angular coordinates of new nodes are significantly smaller than those obtained using randomized
angular coordinates, see Table~\ref{tab:si_table2}, signifying that new connections in these networks are well described by the popularity$\times$similarity optimization.

\begin{table}
\footnotesize
\begin{center}
\begin{tabular}
{|c|c|c|c|}
\hline Network Name & $L$ & $L_{rand}$ & ${\mathcal L}/ { \mathcal L_{rand}}$ \\
\hline Internet (Jan-Apr 2007) & $1100$ & $1800$ & $\exp(700)$ \\
\hline {\it E.coli} metabolic ($S_{0}$--$S_{1}$) & $400$ & $600$ & $\exp(200)$\\
\hline PGP web of trust (Apr-Oct 2003)& $5900$ & $9000$ & $\exp(3100)$\\
\hline
\end{tabular}
\caption{Logarithmic loss calculated only for new-old pairs of nodes.
\label{tab:si_table2}}
\end{center}
\end{table}

\subsection{Example of a network that is not well described by the model}\label{sec:imdb}

We finally present an example of a real network for which our mapping method does not produce good results. Specifically,
we consider the actor network from the Internet Movie Database (IMDb)~\cite{imdb}. To build the network we connect two actors if they
have co-starred in at least one film, limiting our consideration only to films labeled as comedies. The largest connected subgraph
of the resulting network in the year of $2000$ consists of $44936$ actors and has average degree $\bar{k}=13.6$, and average clustering $\bar{c}=0.55$.
This actor network is another example of a growing network with strong clustering and heterogeneous node degrees. However, the mapping of this network using our method
is poor, as illustrated by the connection probability in Fig.~S16.

\begin{figure*}
\centerline{\includegraphics [width=3.0in, height=2in]{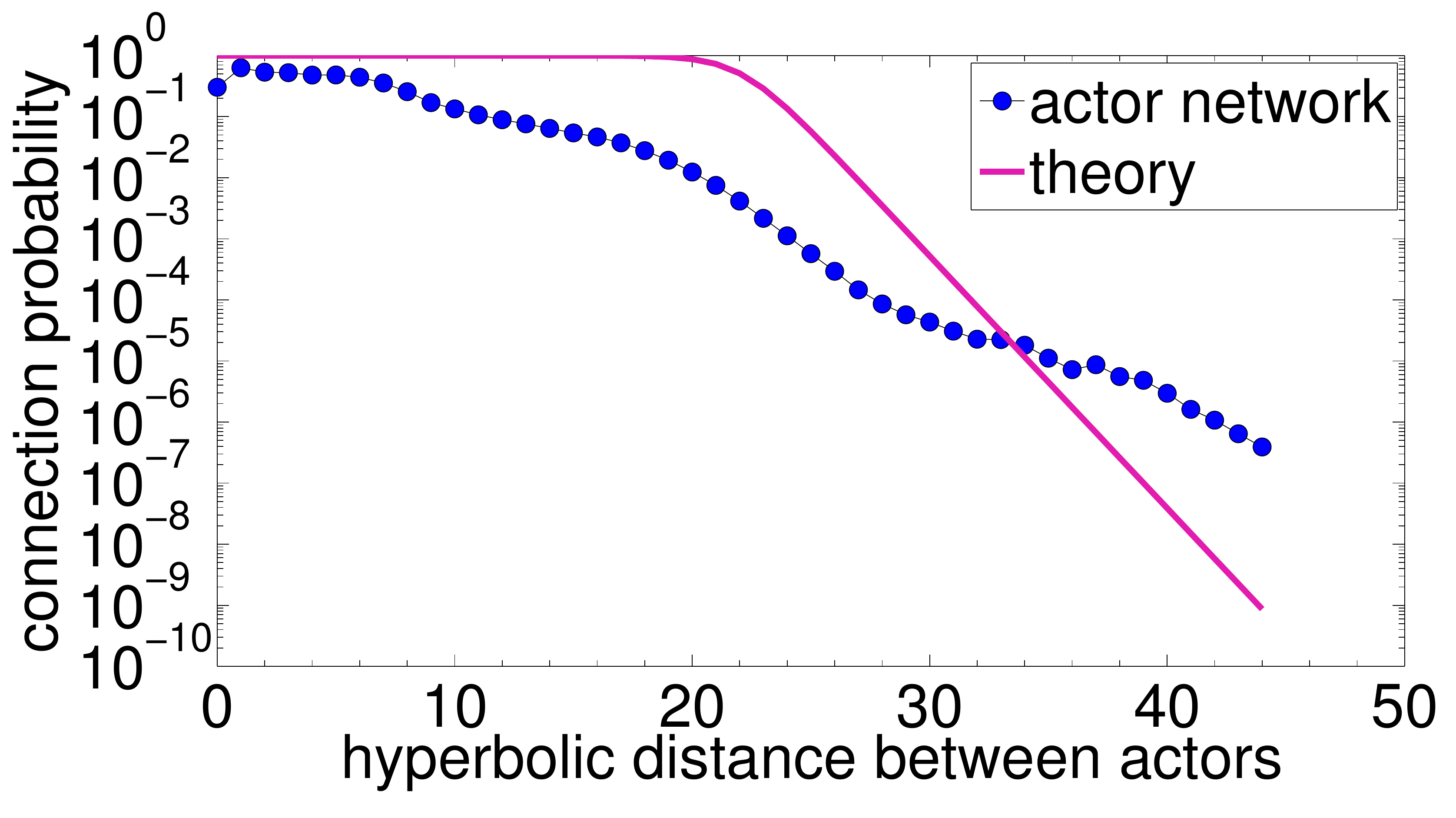}}
\caption{Connection probability for the actor network considered in Section~\ref{sec:imdb}.}
\end{figure*}

We also compute the logarithmic loss for this network using the
inferred node coordinates. The result is $L=6.0\times10^{6}$. This value is larger than the one obtained after randomizing
the node angular coordinates, $L_{rand} = 3.8\times10^{6}$, so that $\mathcal{L} / \mathcal{L}_{rand} =
\exp(L_{rand} - L)=\exp(-2.2\times10^{6})$.

The reason why our model does not describe the actor network well is the following. By construction, the network is overinflated
with fully connected subgraphs, since many modern film crews include hundreds of dissimilar actors. Any pair of such actors participating at least
once in such a large-scale film project, are connected, leading to an abundance of large cliques in the network.
As a result of this overinflation, even not so famous actors that may join the project coming from many different countries, have high chances to be connected.
That is, connections in this network are not well described by popularity$\times$similarity optimization, because even fairly
dissimilar and unpopular actors may be connected with high probability. Therefore, the fact that
this network cannot be successfully mapped by our method is quite expected.

\section{The popularity$\times$similarity model: analysis and simulations}
\label{sec:a1}

In this section we discuss the formulation details of the popularity$\times$similarity model,
analyze its properties, and verify them in simulations.
We start with the simplest version of the model: (1) initially the network is empty;
(2) at time $t \geq 1$, new node $t$ appears having coordinates $(r_t, \theta_t)$,
where $r_t=\ln{t}$, while $\theta_t$ is uniformly distributed on $[0, 2\pi]$, and every
existing node $s$, $s < t$, moves increasing its radial coordinate according to
$r_s(t)=\beta r_s +(1-\beta)r_t$ with parameter $\beta \in [0,1]$; and
(3) node $t$ connects to the $m$ hyperbolically closest nodes $s$, $s < t$; at early times
$t\leq m$, node $t$ connects to all the existing nodes. The value of $m$ controls the average degree in the network $\bar{k}=2m$.
The hyperbolic distance between two points $(r_s, \theta_s)$ and $(r_t, \theta_t)$ is given by~\cite{Bonahon09-book}
\begin{eqnarray}
\label{eq:x_st}
\nonumber x_{st}&=&\frac{1}{2}\,\mathrm{arccosh}\left(\cosh{2 r_s}\cosh{2 r_t}-\sinh{2 r_s} \sinh {2 r_t} \cos{\theta_{st}}\right)\\
\nonumber &\approx& r_s+r_t+\ln(\theta_{st}/2),\quad\textnormal{where}~\theta_{st}=\pi-|\pi-|\theta_s-\theta_t||.
\end{eqnarray}
This expression gives
the distance between two points on the hyperbolic plane of curvature $K=-4$~\cite{Bonahon09-book}.
The model can be generalized for any curvature value (see Section~\ref{sec:zeta_extension}) without
affecting the results since changing the value of curvature
corresponds to simple rescaling of all distances, thus preserving
the distance-induced ordering of nodes, e.g., the sets of $m$ closest nodes, etc.
We call the above model $\textnormal{Model}_1$.

We show in Section~\ref{sec:clustering} that clustering is strongest possible in the networks generated by $\textnormal{Model}_1$. To weaken
clustering we allow connections to nodes farther apart. To do so, we modify step (3) of
$\textnormal{Model}_1$ as follows: (3) new node $t$ picks a randomly chosen  node $s$, $s < t$, and
given that it is not already connected to it, it connects to it with probability $p(x_{st})=1/[1+e^{(x_{st}-R_t)/T}]$,
where parameter $T$ is called network temperature, and $R_t \sim r_t$---the exact value of $R_t$ is specified below.
Node $t$ repeats this step until it gets connected to $m$ nodes. The connection probability $p(x_{st})$ is nothing but the Fermi-Dirac distribution~\cite{KrPa10}.
We call this model $\textnormal{Model}_2$.

We also show in Section~\ref{sec:clustering} that clustering is a decreasing function of temperature, and that at zero temperature we
recover the strongest clustering case, where new nodes connect to the hyperbolically closest
existing nodes. But first we show that for any $\beta \in (0,1)$ both models produce scale-free networks
with the power-law degree distribution identical to the degree distribution in networks growing according to preferential
attachment (PA)~\cite{DoMeSa00}, and having power-law exponent $\gamma=1+ \frac{1}{\beta}$.

\subsection{Degree distribution}
\label{sec:degree_distribution}

We start with $\textnormal{Model}_1$. Consider new node $t$, let $R_t$ be the radius of a hyperbolic
disc centered at this node, and let it connect to all nodes $s$, $s < t$, that lie
within this disc. The probability that there is a connection to node $s$ is
\begin{equation}
P\left[x_{st} \leq R_t\right]=P\left[\theta_{st} \leq 2 e^{-\left(r_s(t)+r_t-R_t\right)}\right]\approx \frac{2}{\pi}e^{-(r_s(t)+r_t-R_t)}.
\label{eq:R_t_prob}
\end{equation}
The average number of existing nodes lying within $R_t$ is
\begin{equation}
\overline{N(R_t)}=\int_{1}^{t} P(x_{it} \leq R_t) di=\frac{2}{\pi} e^{-(r_t-R_t)}\int_{1}^{t}e^{-r_i(t)}di=\frac{2}{\pi} e^{-(r_t-R_t)}\frac{1}{1-\beta}\left(1-e^{-(1-\beta)r_t}\right).
\label{eq:N_R_t}
\end{equation}
Therefore
\begin{equation}
R_t=r_t-\ln\left[\frac{2}{\pi}\frac{\left(1-e^{-(1-\beta)r_t}\right)}{\overline{N(R_t)}(1-\beta)}\right],
\label{eq:R_t}
\end{equation}
is the radius of the hyperbolic disc centered at node $t$, which contains on average the closest
$\overline{N(R_t)}$ existing nodes.  Setting $\overline{N(R_t)}=m$ and substituting $R_t$ from
Equation (\ref{eq:R_t}) into Equation (\ref{eq:R_t_prob}), we  find the probability that an existing
node that appeared at time $s$ attracts a link from a new node $t$, if node $t$ connects on average
to the $m$ closest existing nodes
\begin{equation}
\Pi(r_s(t))=P(x_{st} \leq R_t)=\frac{m}{\frac{1}{1-\beta}\left(1-e^{-(1-\beta)r_t}\right)}e^{-r_s(t)}.
\label{eq:con_prob_1}
\end{equation}
The above equation also holds if the new node $t$ always connects to exactly $m$ closest nodes.
Further, since $\int_{1}^{t}e^{-r_i(t)}di=\frac{1}{1-\beta}\left(1-e^{-(1-\beta)r_t}\right)$,
we can rewrite Equation (\ref{eq:con_prob_1}) as
\begin{equation}
\Pi(r_s(t))=m\frac{e^{-r_s(t)}}{\int_{1}^{t}e^{-r_i(t)}di}=m\frac{e^{-(\beta r_s+(1-\beta)r_t)}}{\int_{1}^{t}e^{-(\beta r_i+(1-\beta)r_t)}di}
=m\frac{\left(\frac{s}{t}\right)^{-\beta}}{\int_{1}^{t}\left(\frac{i}{t}\right)^{-\beta}di}\equiv \Pi_{\textnormal{Model}_1}(s,t).
\label{eq:con_prob_2}
\end{equation}

We now recall how connections are made in PA~\cite{DoMeSa00}, where at sufficiently large times $t$, an existing node $s$
with degree $k_s(t)$ attracts a link from a new node $t$ with probability
\begin{equation}
\Pi(k_s(t))=m\frac{k_s(t)-m+A}{(m+A)t},
\label{eq:con_prob_pa_1}
\end{equation}
where $m$ is the number of existing nodes that each new node connects to,
$A=(\gamma-2)m$ is a parameter called initial attractiveness of each node, and $\gamma$ is the exponent of the target
power law degree distribution. Notice that since each new node brings $m$ connections, at large times $t$ the denominator in
Equation (\ref{eq:con_prob_pa_1}) can be written as
\begin{equation}
(m+A)t = \int_{1}^{t}(\overline{k_i(t)}-m+A)di.
\end{equation}
Further, it has been shown~\cite{DoMeSa00} that
\begin{equation}
\overline{k_s(t)}=m+A\left[\left(\frac{s}{t}\right)^{-\beta}-1\right],
\label{eq:k_s_t}
\end{equation}
where $\beta=\frac{1}{\gamma-1}$, $\beta \in (0,1)$.

The connection probability given by Equation (\ref{eq:con_prob_pa_1}) is conditioned on the exact value of the degree of the node $s$, $k_s(t)$.
Therefore, the unconditional probability that an existing node $s$ attracts a link from a new node $t$,
which can be obtained by Equation (\ref{eq:con_prob_pa_1}) after replacing $k_s(t)$ with its expected value, is
\begin{equation}
\Pi(\overline{k_s(t)})=m\frac{\overline{k_s(t)}-m+A}{\int_{1}^{t}(\overline{k_i(t)}-m+A)di}
=m \frac{\left(\frac{s}{t}\right)^{-\beta}}{\int_{1}^{t} \left(\frac{i}{t}\right)^{-\beta}di} \equiv \Pi_{\textnormal{PA}}(s,t).
\label{eq:con_prob_pa_2}
\end{equation}
From Equations (\ref{eq:con_prob_2}) and (\ref{eq:con_prob_pa_2}) we conclude that
\begin{equation}
\Pi_{\textnormal{Model}_1}(s,t)=\Pi_{\textnormal{PA}}(s,t).
\label{eq:pa_model_equiv}
\end{equation}
This means that for fixed $m$ and $\beta=\frac{1}{\gamma-1}$ the probability that an existing node $s$, $s < t$,
attracts a link from a new node $t$, is the same in $\textnormal{Model}_1$ and PA. This, in turn,
means that the resulting degree distribution in $\textnormal{Model}_1$ is identical to PA, i.e., it is the same power
law with exponent $\gamma=1+\frac{1}{\beta}$, whose exact expression is given by~\cite{DoMeSa00}
\begin{equation}
\label{eq:P_k_theory}
P(k)=(\gamma-1)\frac{\Gamma{\left[(m+1)(\gamma-2)+1\right]}\Gamma{\left[k+m(\gamma-3)\right]}}{\Gamma{\left[m(\gamma-2)\right]}\Gamma{\left[k+m(\gamma-3)+\gamma\right]}}.
\end{equation}
Further, knowing the current degree of a node $k$, the node attracts a link from a new node $t$ with probability as in Equation (\ref{eq:con_prob_pa_1})
\begin{equation}
\label{eq:pi_k_theory}
\Pi(k)=m\frac{k-m+A}{(m+A)t}.
\end{equation}
Probabilities $P(k)$ and $\Pi(k)$ are both defined for $k \geq m$.
Finally, using Equation (\ref{eq:k_s_t}), we can deduce that
\begin{equation}
\overline{k_s(t)}=m+A\left[e^{-(r_s(t)-r_t)}-1\right]\sim e^{-(r_s(t)-r_t)}.
\label{eq:k_r}
\end{equation}

In contrast to PA where the case $\gamma=2$ is problematic~\cite{DoMeSa00},
there are no problems with $\gamma=2$ in $\textnormal{Model}_1$, where $\gamma=2$ corresponds to $\beta=1$, i.e., to the case where nodes do not move.
It is easy to check that for $\beta \rightarrow 1$, $\int_{1}^{t}e^{-r_i(t)}di=\frac{1}{1-\beta}\left(1-e^{-(1-\beta)r_t}\right)\rightarrow r_t$, and Equations (\ref{eq:N_R_t}), (\ref{eq:R_t}), and (\ref{eq:con_prob_1}) are all well defined.

We now move to $\textnormal{Model}_2$, and show that the same results with respect to the degree
distribution hold there as well. Recall that in $\textnormal{Model}_2$ a new node $t$, instead of connecting
to the $m$ closest nodes, picks a random existing node $s$, $s < t$, and
given that it is not already connected to it, it
connects to it with probability $p(x_{st})=1/[1+e^{(x_{st}-R_t)/T}]$. It then repeats this
procedure until it gets connected to $m$ nodes. Notice that at long times $t \gg m$, the probability that
node $t$ selects a random node $s$ to which it is already connected, is insignificant and can be ignored to ease
analysis. Further, notice that the probability $p(x_{st})$ can be also written as
\begin{equation}
\label{eq:p_x_st}
p(x_{st})=\frac{1}{1+\left(X(s,t)\frac{\theta_{st}}{2}\right)^{\frac{1}{T}}}, \quad\textnormal{where~} X(s,t)=e^{(r_s(t)+r_t-R_t)}.
\end{equation}
Since node $t$ picks a random existing node and $\theta_{st}$ is uniformly distributed in $[0,\pi]$,
the probability that node $t$ connects to node $s$ is
\begin{equation}
P(s,t)=\frac{1}{t}\frac{1}{\pi} \int_{0}^{\pi}\frac{1}{1+\left(X(s,t)\frac{\theta_{st}}{2}\right)^{\frac{1}{T}}} d\theta_{st} \approx\frac{2T}{t\sin{T\pi}}\frac{1}{X(s,t)}.
\label{eq:P_s_t}
\end{equation}
The approximation in Equation (\ref{eq:P_s_t}) holds for $T < 1$. Now, the probability that node $t$ connects to any node is
\begin{equation}
P(t)=\int_{1}^{t}P(i,t)di.
\label{eq:P_t}
\end{equation}
Since node $t$ brings $m$ new links, then at sufficiently large times $t$, the probability that node $s$ attracts a link is
\begin{equation}
\Pi_{\textnormal{Model}_2}(s,t)=m\frac{P(s,t)}{P(t)}=m\frac{e^{-r_s(t)}}{\int_{1}^{t}e^{-r_i(t)}di}=\Pi_{\textnormal{Model}_1}(s,t)=\Pi_{\textnormal{PA}}(s,t).
\label{eq:con_prob_model_2}
\end{equation}
This means that for fixed $m$ and $\beta=\frac{1}{\gamma-1}$, the degree distribution and link attraction probability in $\textnormal{Model}_2$ are the same as in $\textnormal{Model}_1$, i.e., given by Equations~(\ref{eq:P_k_theory}) and~(\ref{eq:pi_k_theory}). The limit $\beta \rightarrow 1$ is also well defined.

Notice that as $T \rightarrow 0$, $p(x_{st}) \rightarrow 1$ if $x_{st} \leq R_t$, and $p(x_{st}) \rightarrow 0$ if $x_{st} > R_t$.
In this case, setting $R_t$ as in Equation (\ref{eq:R_t}) with $\overline{N(R_t)}=m$, constrains the connections of
a new node $t$ to its $m$ hyperbolically closest nodes, and $\textnormal{Model}_2$ becomes identical to $\textnormal{Model}_1$.
In $\textnormal{Model}_2$, we can also compute the average number of existing nodes lying within $R_t$ from a new node $t$
\begin{equation}
\label{eq:N_R_t_T}
\overline{N(R_t)}=t P(t)= \frac{2T}{\sin{T\pi}} e^{-(r_t-R_t)}\frac{1}{1-\beta}\left(1-e^{-(1-\beta)r_t}\right).
\end{equation}
Therefore, in analogy to $\textnormal{Model}_1$, setting $\overline{N(R_t)}=m$ we can fix $R_t$
\begin{equation}
R_t=r_t-\ln\left[\frac{2T}{\sin{T\pi}}\frac{\left(1-e^{-(1-\beta)r_t}\right)}{m (1-\beta)}\right].
\label{eq:R_t_T}
\end{equation}
Equation (\ref{eq:R_t_T}) is valid for $0 < T < 1$, and for $T \rightarrow 0$ it becomes Equation
(\ref{eq:R_t}) as expected.

Figure~S2 shows simulation results for $\textnormal{Model}_2$, and Fig.~2(a) with Fig.~S3(a) show simulation results
for $\textnormal{Model}_1$, validating our analysis. Figure~S3(b) also shows that clustering is strong in networks growing according
to popularity$\times$similarity optimization, as opposed to PA. We study clustering in the next section.

\setcounter{figure}{1}
\begin{figure*}
\centerline{\includegraphics[width=3in]{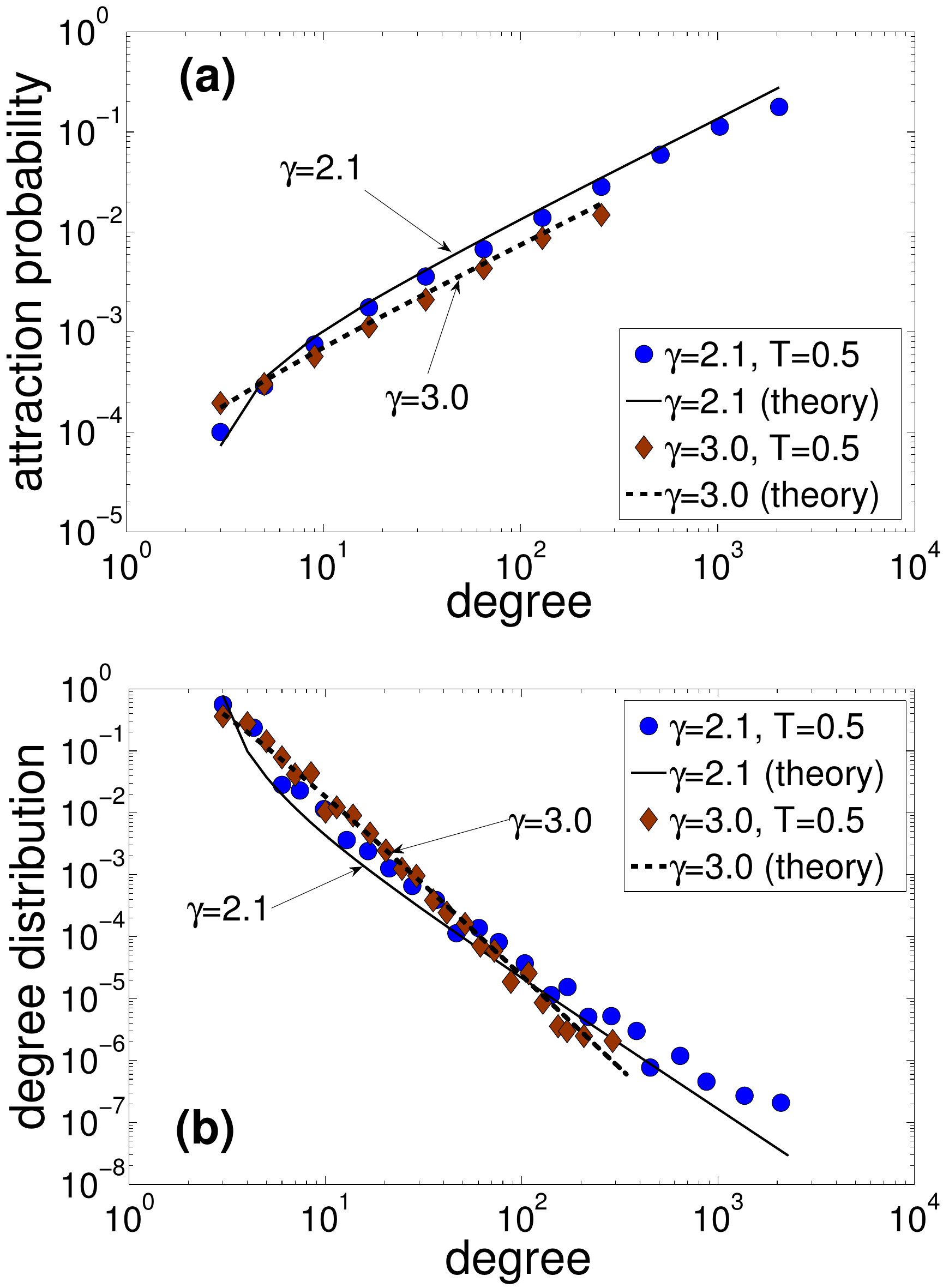}}
\caption{Plot (a) shows the probability $\Pi(k)$ that an existing node of degree $k$ attracts a link in networks grown up to  $t=10^4$ nodes according
to $\textnormal{Model}_{2}$, with $T=0.5$, $m=3$, and $\gamma=2.1, 3.0$. The plot also shows the corresponding theoretical predictions
given by Equation (\ref{eq:pi_k_theory}). Plot (b) shows the distribution $P(k)$ of node degrees in the same networks. The theoretical predictions are given
by Equation  (\ref{eq:P_k_theory}). Small deviations of the theoretical prediction for $\gamma=2.1$ are due to the increasingly pronounced
finite-size effects at $\gamma\to2$~\cite{BoPaVe04}. Similar results hold for other values of $\gamma \geq 2$,  $0 \leq T < 1$, and $m$, not shown to avoid clutter.}
\end{figure*}

\begin{figure*}
\centerline{\includegraphics[width=3in]{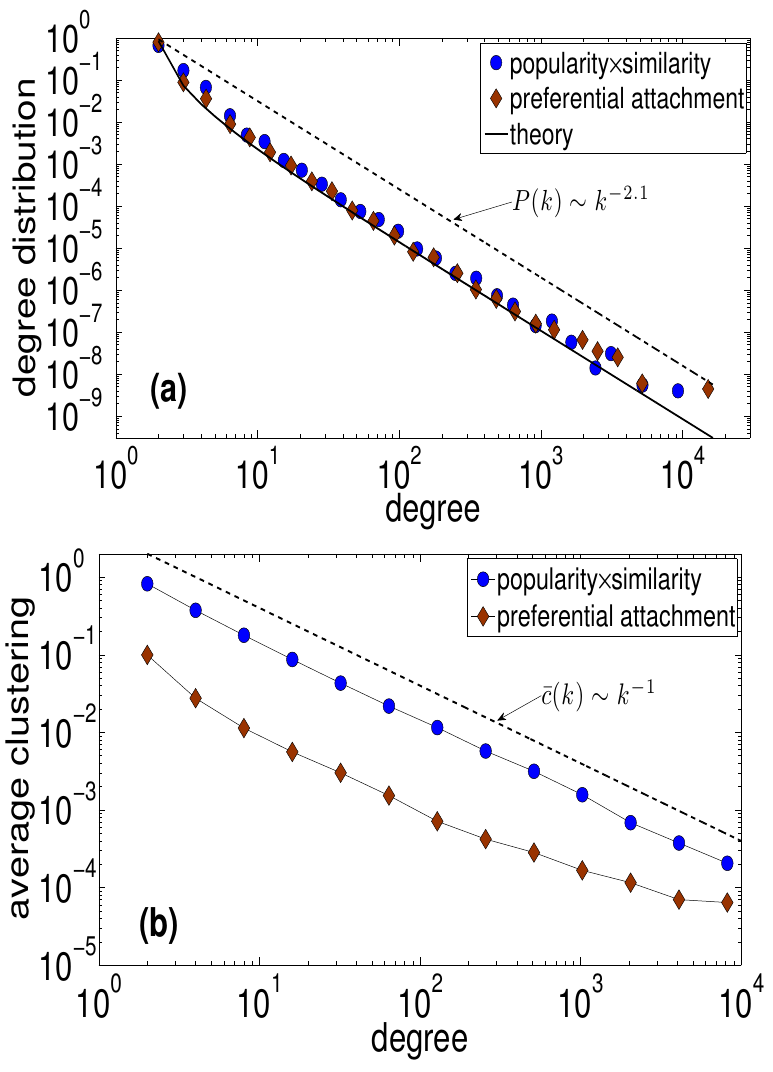}}
\caption{Plot (a) shows the distribution $P(k)$ of node degrees for the two networks considered in Fig.~2.
Plot~(b) shows for the same two networks the average clustering $\bar{c}(k)$ of $k$-degree nodes, defined as the ratio of the number of triangles involving a $k$-degree node to the maximum such number $k(k-1)/2$, averaged over all the $k$-degree nodes. The $1/k$ scaling of $\bar{c}(k)$ is often considered
as a signature of the network's hierarchical organization~\cite{RaSoMoOlBa02}. The average clustering $\bar{c}=\sum_k\bar{c}(k)P(k)$ in
the optimization and PA networks is $\bar{c}=0.83$ and $\bar{c}=0.12$, respectively, as mentioned in Fig.~2.}
\end{figure*}

\subsection{Clustering}
\label{sec:clustering}

We have shown that networks grown according to popularity$\times$similarity optimization have an effective hyperbolic
geometry underneath, from which power-law degree distributions emerge. We now show that the metric property of
this geometry, i.e., the triangle inequality, leads to strong clustering in these networks, i.e., the large number
of triangular subgraphs.

Intuitively, if node $a$ is hyperbolically close to a node $b$, and $b$ is close to a third node $c$,
then $a$ is also close to $c$ because of the triangle inequality. Since all three nodes are close to each other, links between all of them
forming triangle $abc$ exist with high probability. This probability depends on the value of the temperature $T \in [0,1)$.

We show next that average clustering at time $t$, $\bar{c}(t)$, is a decreasing function of temperature: clustering
is maximized at $T=0$, and it gradually decreases to zero as $T \rightarrow 1$.

\subsubsection{Analysis}

Let $\bar{c}(s,t)$ be the average clustering of node $s$ at time $t$. Then
\begin{equation}
\label{eq:bar_c_t}
\bar{c}(t)=\frac{1}{t}\int_{1}^{t}\bar{c}(s,t) ds.
\end{equation}
where $\bar{c}(s,t)$ is given by~\cite{BoPa03}
\begin{equation}
\label{eq:bar_c_s_t}
\bar{c}(s,t)=\frac{2 \overline{T_s(t)}}{[\overline{k_s(t)}]^2},
\end{equation}
and $\overline{T_s(t)}$ is the expected number of triangles
that contain node $s$ at time $t$, while $\overline{k_s(t)}$ is $s$'es expected degree given by Equation (\ref{eq:k_s_t}).
To compute $\overline{T_s(t)}$ we break it into two parts: (i) $\overline{T_s^\textnormal{old}}$, which is the expected number of triangles
formed when node $s$ appeared, i.e., by connections from node $s$ to existing pairs of connected nodes; and (ii)
$\overline{T_s^\textnormal{new}(t)}$, which is the expected number of triangles formed by new nodes appearing after node $s$, i.e.,
by connections from new nodes to old pairs of connected nodes where one of the nodes is node $s$. Clearly, $\overline{T_s(t)}=\overline{T_s^\textnormal{old}}+\overline{T_s^\textnormal{new}(t)}$.

The probability that two nodes $s < t$ are connected in $\textnormal{Model}_2$ given the hyperbolic distance $x_{st}$
between them, is $m\frac{\frac{1}{t}p(x_{st})}{P(t)}=m\frac{\frac{1}{t}p(x_{st})}{\frac{m}{t}}=p(x_{st})$, i.e., they connect
with probability given by Equation (\ref{eq:p_x_st}). Introducing notation $\chi_{st}=X(s,t)\frac{\theta_{st}}{2}$, we can write
\begin{equation}
\label{eq:c_v_1}
p(x_{st})=\frac{1}{1+\chi_{st}^{\frac{1}{T}}}=\widetilde{p}(\chi_{st}).
\end{equation}
Since $T < 1$, the function $\widetilde{p}(\chi)$ is integrable
\begin{equation}
\label{eq:c_v_2}
I=\int_{0}^{\infty}\frac{1}{1+\chi^\frac{1}{T}}d\chi=\frac{T\pi}{\sin{T\pi}}.
\end{equation}
Further, since $X(s,t)=e^{(r_s(t)+r_t-R_t)}$, with $R_t$ given by Equation (\ref{eq:R_t_T}) we can also write
\begin{equation}
X(s,t)= \frac{2T}{\sin{T\pi}}f(s,t),\quad\textnormal{where}~f(s,t)=\frac{1-e^{-(1-\beta)r_t}}{m(1-\beta)}e^{r_s(t)} = \frac{t^{1-\beta}-1}{m(1-\beta)s^{-\beta}}.
\end{equation}
Now, the probability that three nodes $s, t', t'' < t$ form a triangle, is the probability that the three nodes are connected.
Let $\theta_{t'}, \theta_{t''}$ be the angular coordinates of nodes $t'$ and $t''$ respectively, and $\theta_s$ be the angular coordinate of node $s$.
As the angular coordinate is uniformly distributed, we can set without loss of generality $\theta_s=0$. Therefore, with $\theta_{t's}=\theta_{st'}=\theta_{t'}, \theta_{t''s}=\theta_{st''}=\theta_{t''}$, and
$\theta_{t't''}=\theta_{t''t'}=\theta_{t'}-\theta_{t''}$, it is easy to see that
{\begin{eqnarray}
\overline{T_s^\textnormal{old}}&=& \frac{1}{4\pi^2}\int_{1}^{s}dt'
\int_{1}^{t'}dt'' \int_{-\pi}^{\pi} \int_{-\pi}^{\pi}d\theta_{t'} d\theta_{t''}\widetilde{p}(|\chi_{t's}|)\widetilde{p}(|\chi_{t''t'}|)\widetilde{p}(|\chi_{t''s}|).\nonumber\\
\overline{T_s^\textnormal{new}(t)}&=&\frac{1}{4\pi^2} \int_{s}^{t}dt'\left\{\int_{1}^{s}dt''\int_{-\pi}^{\pi} \int_{-\pi}^{\pi}d\theta_{t'} d\theta_{t''} \widetilde{p}(|\chi_{st'}|) \widetilde{p}(|\chi_{t''t'}|)\widetilde{p}(|\chi_{t''s}|)\right.\nonumber\\
&+&\left.\int_{s}^{t'}dt''\int_{-\pi}^{\pi} \int_{-\pi}^{\pi}d\theta_{t'} d\theta_{t''} \widetilde{p}(|\chi_{st'}|) \widetilde{p}(|\chi_{t''t'}|)\widetilde{p}(|\chi_{st''}|)\right\}.\label{eq:T_s_t_1}
\end{eqnarray}}
Changing the $\theta$ integration variables in Equation (\ref{eq:T_s_t_1}) to the corresponding $\chi$ variables gives
\begin{eqnarray}
\overline{T_s^\textnormal{old}}&=& \frac{1}{(2I)^2} \int_{1}^{s}\frac{dt'}{f(t',s)}\int_{1}^{t'}\frac{dt''}{f(t'',s)}\nonumber\\
&\times& \int_{-If(t',s)}^{I f(t',s)}d\chi'\int_{-I f(t'',s)}^{I f(t'',s)}d\chi''\widetilde{p}(|\chi'|)\widetilde{p}\left(f(t'',t')\left|\frac{\chi'}{f(t',s)}-\frac{\chi''}{f(t'',s)}\right|\right)\widetilde{p}(|\chi''|).\nonumber\\
\overline{T_s^\textnormal{new}(t)} &=& \frac{1}{(2I)^2} \int_{s}^{t}\frac{dt'}{f(s, t')}\nonumber\\
&\times& \left\{\int_{1}^{s}\frac{dt''}{f(t'',s)} \int_{-I f(s, t')}^{I f(s,t')}d\chi' \int_{-I f(t'',s)}^{I f(t'',s)} d\chi''\widetilde{p}(|\chi'|)\widetilde{p}\left(f(t'',t')\left|\frac{\chi'}{f(s, t')}-\frac{\chi''}{f(t'',s)}\right|\right)\widetilde{p}(|\chi''|)\right.\nonumber\\
&+& \left. \int_{s}^{t'} \frac{dt''}{f(s, t'')} \int_{-I f(s, t')}^{I f(s,t')}d\chi' \int_{-I f(s, t'')}^{I f(s, t'')} d\chi''\widetilde{p}(|\chi'|)\widetilde{p}\left(f(t'',t')\left|\frac{\chi'}{f(s, t')}-\frac{\chi''}{f(s, t'')}\right|\right) \widetilde{p}(|\chi''|)\right\},\nonumber\\
\label{eq:T_s_t_2}
\end{eqnarray}
which cannot be written as a closed-form expression. However, these equations allow us to infer the relationship
between the clustering strength of the network and parameter $T$. As $T \rightarrow 1$, $I \rightarrow \infty$,
and therefore, $\overline{T_s^\textnormal{old}}\rightarrow 0$, $\overline{T_s^\textnormal{new}(t)} \rightarrow 0$, $\forall s,t$,
meaning that clustering goes to zero. As $T \rightarrow 0$, $I \rightarrow 1$, and clustering is maximized. To see this, consider
the node with the smallest degree, i.e., the node that appeared at time $s=t$, whose degree is $k_{t}(t)=m$. Clearly, $\overline{T_t^\textnormal{new}(t)}=0$. To compute
$\overline{T_t^\textnormal{old}}$, observe that when $T \rightarrow 0$, $\widetilde{p}(\chi) \rightarrow \Theta(1-\chi)$, and therefore, the inner
integrals taken over the variables $\chi', \chi''$, reduce to the area of intersection of the square defined by $\{|\chi'| < 1; |\chi''| <1\}$, and the stripe
$f(t'',t')\left|\frac{\chi'}{f(t',t)}-\frac{\chi''}{f(t'',t)}\right| < 1$. For most of the combinations of $t', t''$ the stripe is so wide that
it fully contains the square whose area is 4, yielding at large $t$, $\overline{T_t^\textnormal{old}} \approx \frac{m^2}{2}$. Given Equation~(\ref{eq:bar_c_s_t}),
this means that $\bar{c}(t,t) \approx 1$, proving that clustering is maximized at the zero temperature. Recall that clustering cannot be equal to its maximum possible value of $1$ for all node degrees because of structural constraints imposed by power-law degree distributions~\cite{SeBo05}.
For arbitrary values of $s < t$ we need to compute $\overline{T_s^\textnormal{new}(t)}$, but the inner integration region
defined by the $\chi', \chi''$ variables in the expression for $\overline{T_s^\textnormal{new}(t)}$~(\ref{eq:T_s_t_2})
depends on the exact mutual relationship between $s$, $t'$, and $t''$, making
the analytic computation unfeasible. However, one can check that $\bar{c}(s,t)$ increases as $s$ increases, and that
average clustering decreases almost linearly with $T \in (0,1)$.

\subsubsection{Simulations}

Figure~S4 shows average clustering in simulated networks. As predicted by our analysis, clustering decreases as $T$ increases,
and vanishes as $T$ approaches $1$. Clustering is also the stronger, the smaller the $\gamma$.

\begin{figure*}
\centerline{\includegraphics [width=3in]{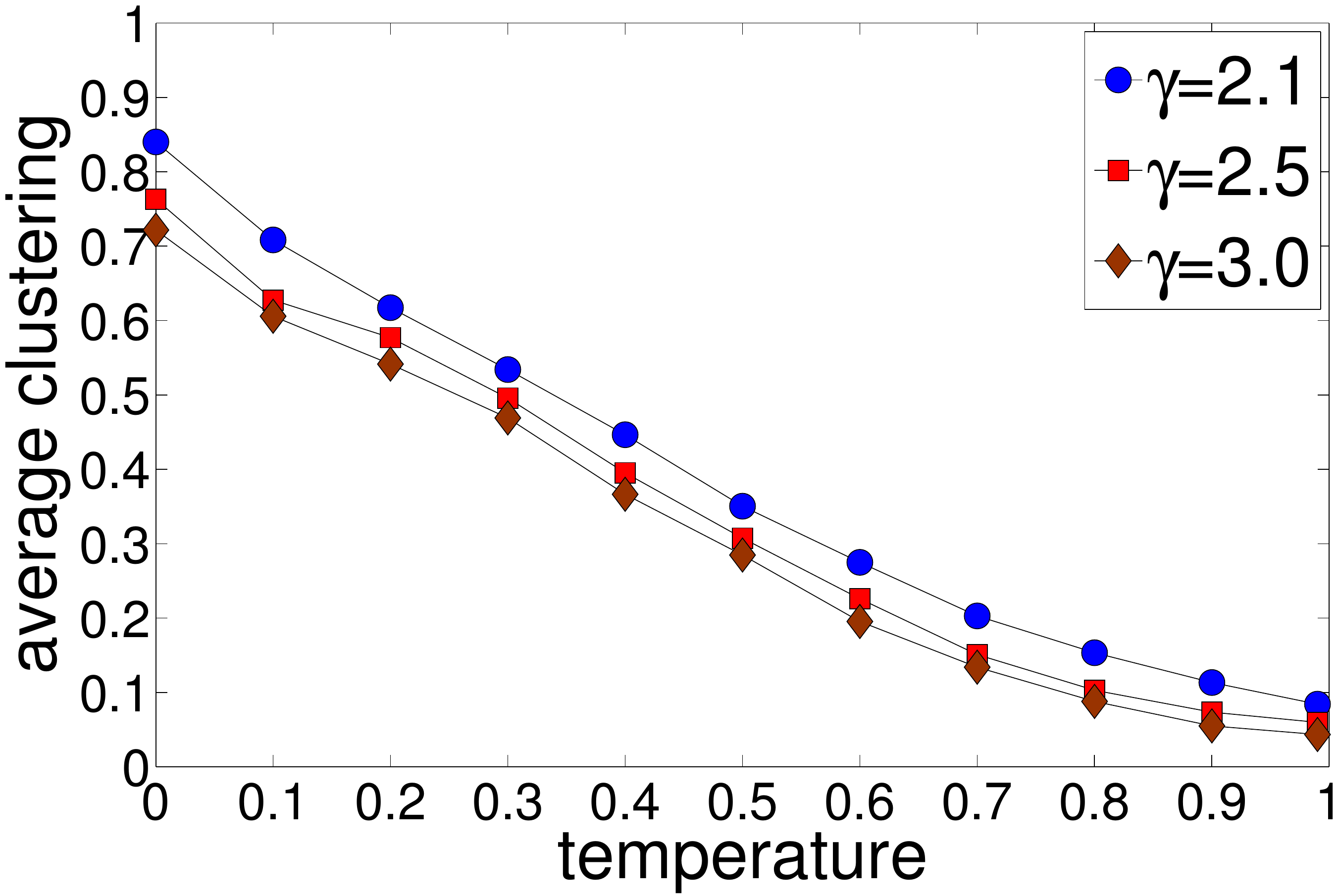}}
\caption{Average clustering $\bar{c}(t)$ at $t=10^4$ as a function of temperature $T\in[0,1)$ in networks grown according to popularity$\times$similarity optimization with $m=3$.}
\end{figure*}

To confirm that zero temperature yields the strongest \emph{possible} clustering (modulo fluctuations),
we perform the following experiment. We grow three networks up to $t=1000$ nodes according to $\textnormal{Model}_1$
with $\gamma=2.1, 2.5, 3.0$ and $m=3$. The average clustering in these networks is $\bar{c}=0.83, 0.76, 0.72$, respectively.
For each network we then perform a number of random link rewirings preserving the degree distribution in the network and trying to increase its clustering if possible~\cite{MaSneZa04}. Specifically, we select a random pair
of links A--B and C--D in the network, and rewire them to A--D and B--C, provided that none of these
links already exist in the network and that the rewiring will not decrease clustering. If these two conditions
are met, then the rewiring is accepted, otherwise it is aborted, and a new pair of links is selected. This way each accepted rewiring step
preserves the degree distribution in the network, and can only increase its average clustering.
For each network we run the experiment until $2000$ rewiring steps were accepted, measuring the new average clustering $\bar{c}_{new}$ every $100$ accepted rewirings.
Figure~S5 shows the results. From the figure, we observe only a minor increase of clustering from its original value, quickly reaching saturation as the number of accepted rewirings increases, as expected. After $2000$ accepted rewirings the average clustering is $\bar{c}_{new}=0.86, 0.81, 0.78$ for $\gamma=2.1, 2.5, 3.0$.

\begin{figure*}
\centerline{\includegraphics [width=3in]{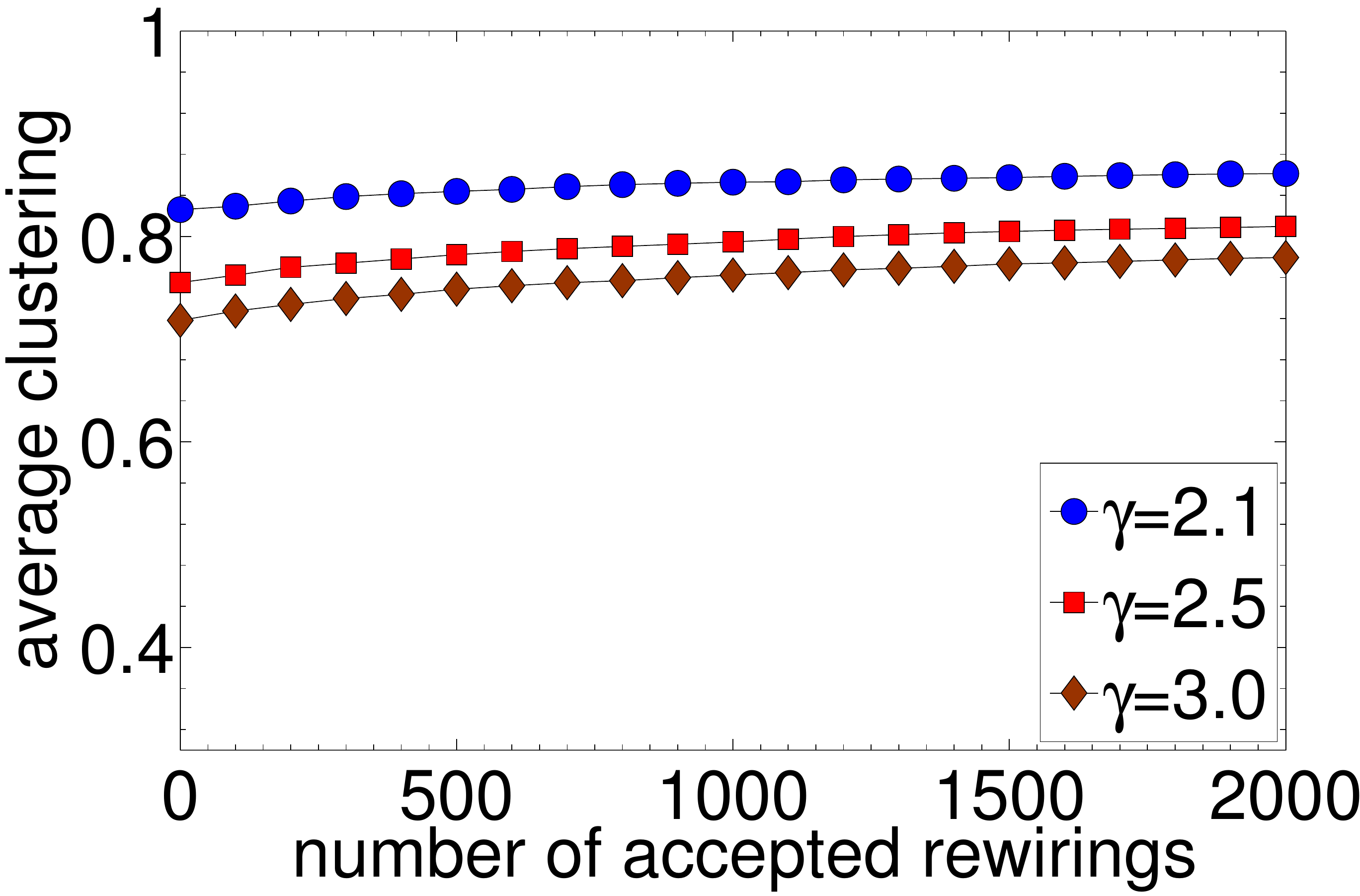}}
\caption{Average clustering as a function of the number of accepted clustering-increasing rewirings in networks grown according to $\textnormal{Model}_{1}$ with $m=3$.}
\end{figure*}

\subsection{Connecting to nodes within distance $R_t$, and densification}

We now consider a variant of the popularity$\times$similarity model, where a new node $t$, instead of connecting to
\emph{exactly} $m$ existing nodes as in $\textnormal{Model}_2$, looks at every existing node $s$, $s <t$, only once and connects to it
with probability $p(x_{st})$ given by Equation (\ref{eq:p_x_st}). We call this variant $\textnormal{Model}_{2'}$.
In this case, the probability that node $s$ attracts a link from node $t$ is $\Pi_{\textnormal{Model}_{2'}}(s,t)=t P(s,t)$,
where $P(s,t)$ as given by Equation (\ref{eq:P_s_t}). The average number of nodes that node $t$ connects to
is $\overline{N(R_t)}=\int_{1}^{t} \Pi_{\textnormal{Model}_{2'}}(i,t) di =t \int_{1}^{t}P(i,t) di = t P(t)$,
with $P(t)$ given by Equation (\ref{eq:P_t}). That is, $\overline{N(R_t)}$ is given again by Equation (\ref{eq:N_R_t_T}) and
can be fixed to $m$ by setting $R_t$ as in Equation (\ref{eq:R_t_T}). Further, since $t=\frac{\overline{N(R_t)}}{P(t)}=\frac{m}{P(t)}$, we have
\begin{equation}
\Pi_{\textnormal{Model}_{2'}} (s,t) =m\frac{P(s,t)}{P(t)}=m\frac{e^{-r_s(t)}}{\int_{1}^{t}e^{-r_i(t)}di}.
\label{eq:con_prob_model_2'}
\end{equation}
That is, $\textnormal{Model}_{2'}$ is equivalent to $\textnormal{Model}_{2}$ (cf.~Eq.~(\ref{eq:con_prob_model_2})) with the difference that in $\textnormal{Model}_{2'}$
a new node $t$ connects on \emph{average} to $m$ existing nodes.

Parameter $T \in [0,1)$ can be used again to tune clustering. As $T \rightarrow 0$ a new node $t$ connects only to all nodes
within distance $R_t$ from it, and we have a variant of $\textnormal{Model}_1$ where clustering is maximized. Indeed, in this case,
the probability that node $s$ attracts a link from a new node $t$ is given again by Equation (\ref{eq:R_t_prob}), which means
that Equations (\ref{eq:N_R_t}), (\ref{eq:R_t}), (\ref{eq:con_prob_1}) and (\ref{eq:con_prob_2}) hold here as well.
The difference here is that the new node $t$ connects to closest nodes whose \emph{average} number is $m$.

To quantify the difference between $\textnormal{Model}_2$ and $\textnormal{Model}_{2'}$, we need to consider
the distribution of the number of existing nodes that a new node $t$ connects to in $\textnormal{Model}_{2'}$,
and to check how narrowly distributed this number is around its average value $m$. The connection events
are statistically independent, so that the number of connections to existing nodes is a sum of independent Bernoulli
trials with different success probabilities $\Pi_{\textnormal{Model}_{2'}}(s,t)$. Hence, the distribution of $N(R_t)$ follows
the Poisson-Binomial distribution with average $m$ and variance $\sigma^2 (t) \approx \int_{1}^{t}\left(1-\Pi_{\textnormal{Model}_{2'}}(i,t)\right)\Pi_{\textnormal{Model}_{2'}}(i,t)di$.
We do not use strict equality in the formula for $\sigma^2(t)$ as we replace the summation with the integration to ease the calculations. Performing the integration we can see that
\begin{equation}
\sigma^2(t) \approx m-g(m, \beta, t),
\end{equation}
where $g(m, \beta, t)$ a function of $m$, $\beta$, and $t$ that goes to zero as $t \rightarrow \infty$. Therefore at $t \rightarrow \infty$ the variance $\sigma^2(t)$ approaches $m$, which is the variance of a Poisson distribution with the average equal to $m$. Indeed, by Le Cam's Theorem \cite{LeCam60}
$\sum_{i=0}^{\infty}|P(N(R_t)=i)-\frac{\lambda^i e^{-\lambda}}{i!}| < 2 \int_{1}^{t}\left(\Pi_{\textnormal{Model}_{2'}}(i,t)\right)^2 di \rightarrow 0$ at $t \rightarrow \infty$, and
therefore the distribution of $N(R_t)$ converges to the Poisson distribution with the average at $m$.

The simulation results in Fig.~S6 confirm the analysis above. Figure~S7 shows the simulation results for the average clustering as a function of temperature, where the behavior is similar to Fig.~S4 as expected. Finally, in Fig.~S8 we repeat the same experiment with the same parameter values as in Fig.~S5, verifying that networks grown according to $\textnormal{Model}_{2'}$ with $T=0$ have maximum possible clustering.

\begin{figure*}
\centerline{\includegraphics[width=3in]{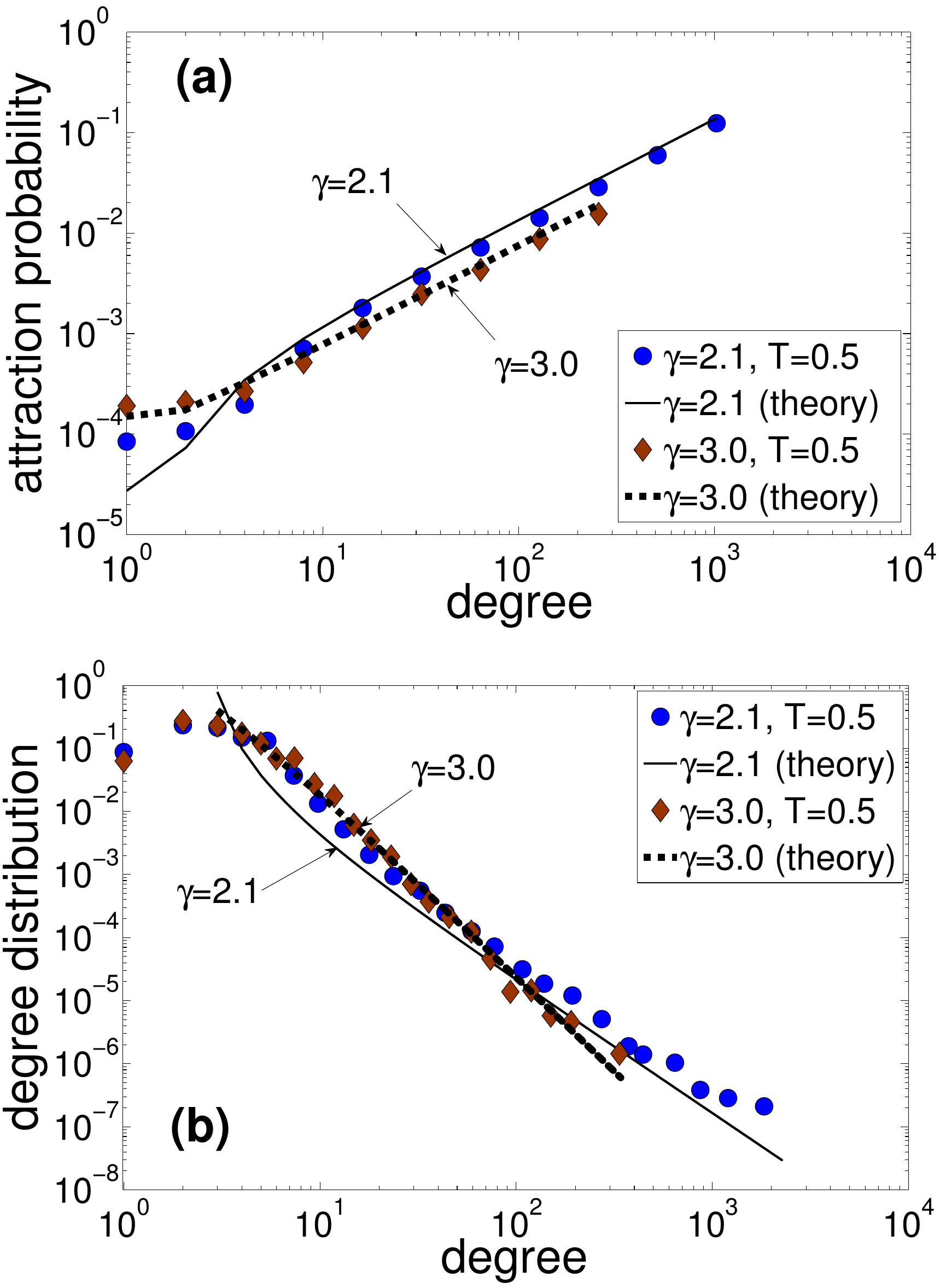}}
\caption{Plot (a) shows the probability $\Pi(k)$ that an existing node of degree $k$ attracts a link in networks grown up to  $t=10^4$ nodes according
to $\textnormal{Model}_{2'}$, with $T=0.5$ and $\gamma=2.1, 3.0$. Each new node connects on average to $m=3$ existing nodes. The theoretical predictions
are given by Equation (\ref{eq:pi_k_theory}) when $ k \geq m$, and  when $k < m$  are given by the formula $\Pi(k)=m\frac{A}{(m+A)t}$.
Plot (b) shows the distribution $P(k)$ of node degrees in the same networks. The theoretical predictions are given
by Equation  (\ref{eq:P_k_theory}), which is defined only for $k \geq m$. Compared to Fig.~S2, here
we observe stronger deviations of the distributions from the power laws at small degrees $k$,
due to fluctuations of the initial degree of a node around its average value $m=3$.
Similar results hold for other values of $\gamma \geq 2$,  $0 \leq T < 1$, and $m$,
not shown to avoid clutter.}
\end{figure*}

\begin{figure*}
\centerline{\includegraphics [width=3in]{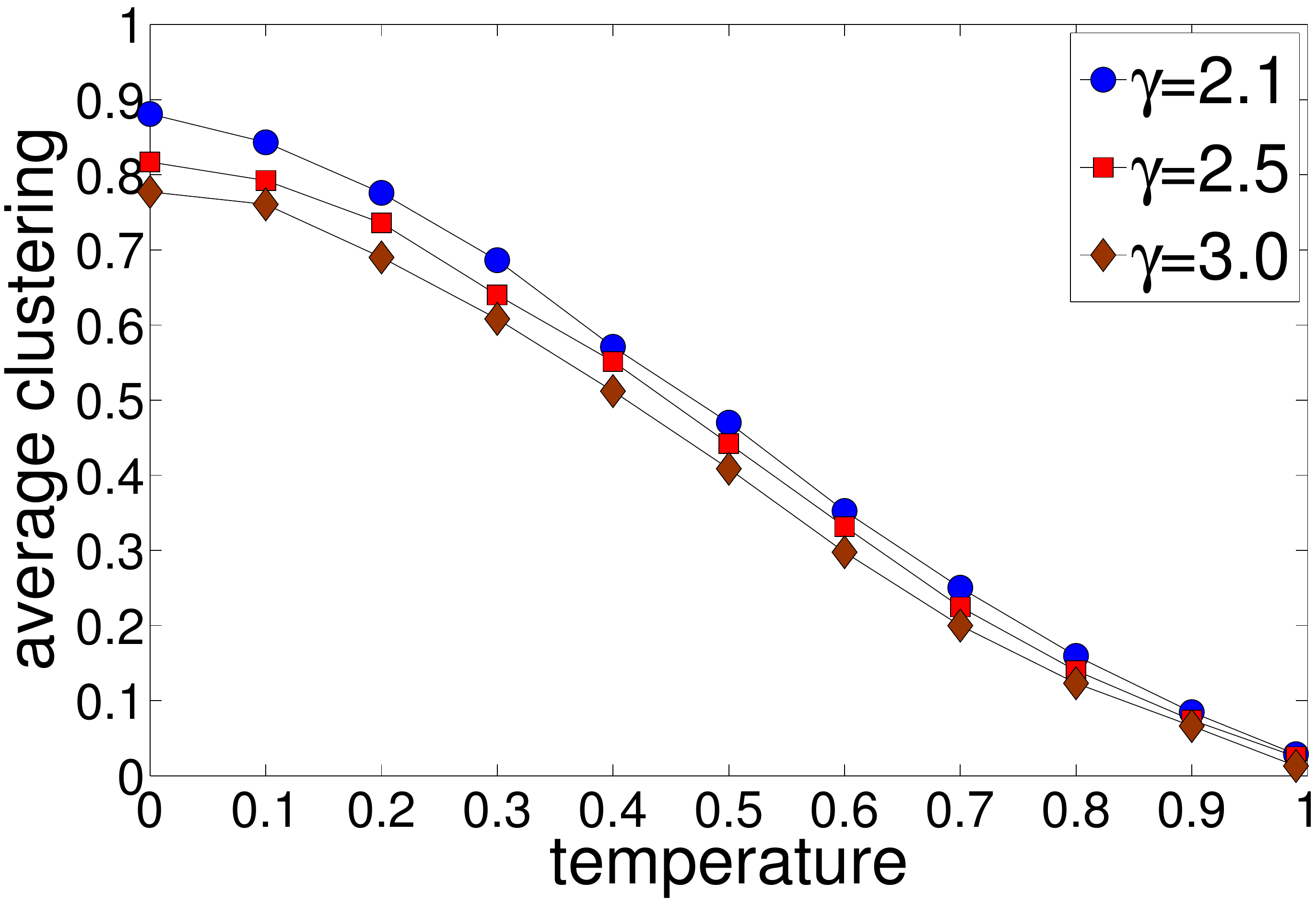}}
\caption{Average clustering $\bar{c}(t)$ at $t=10^4$ as a function of temperature $T\in[0,1)$ in networks growing according to $\textnormal{Model}_{2'}$,  where each new node connects on average to $m=3$ existing nodes. Clustering is calculated excluding nodes of degree $1$, whose clustering is always zero.}
\end{figure*}

\begin{figure*}
\centerline{\includegraphics [width=3in]{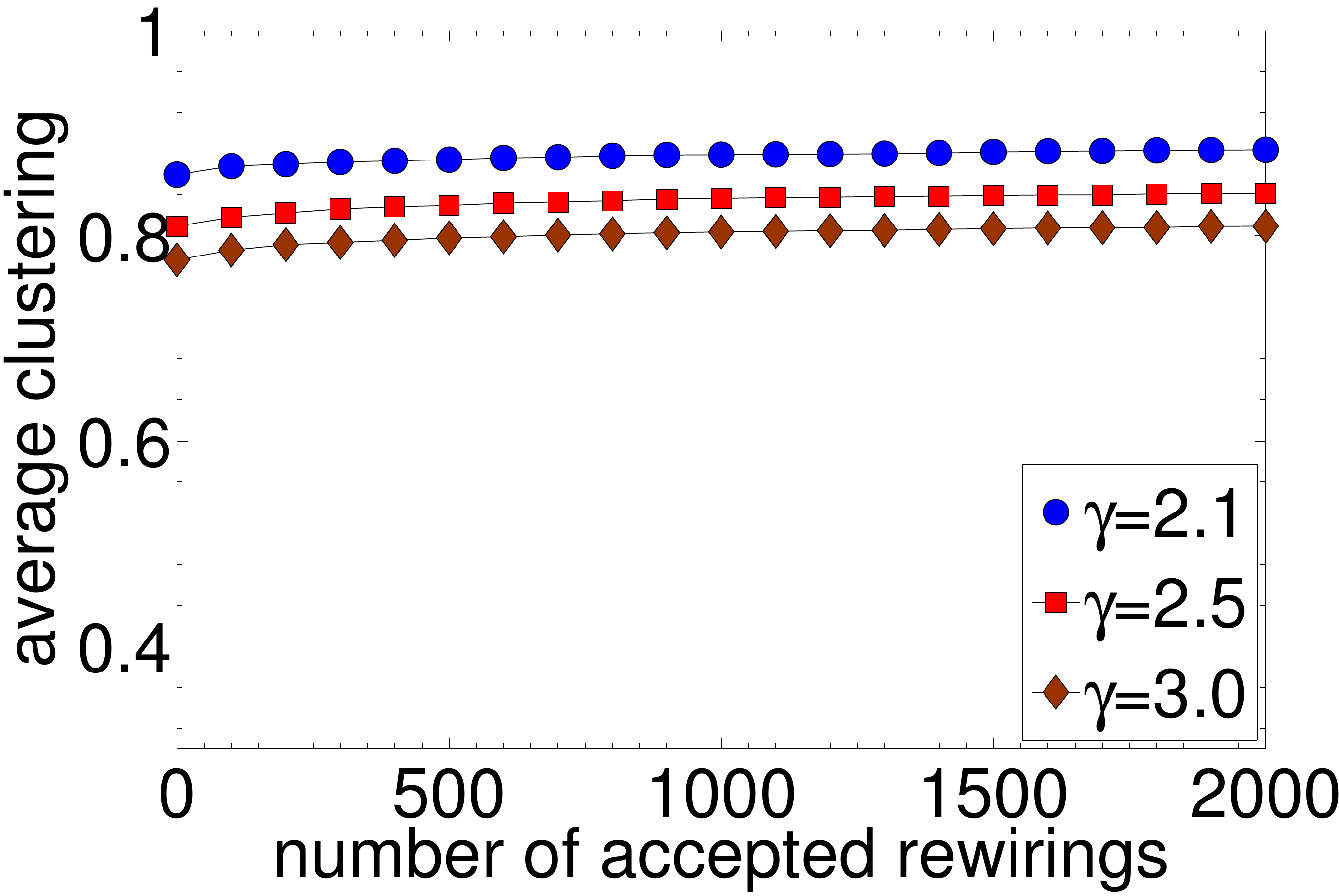}}
\caption{Average clustering as a function of the number of accepted clustering-increasing rewirings for networks grown according to $\textnormal{Model}_{2'}$ with $T=0$ and $m=3$.}
\end{figure*}

Finally, if the connection disc radius is $R_t=r_t$ instead of Equation (\ref{eq:R_t_T}), then the average degree is not constant $\bar{k}=2m$, but grows with the network size $t$, an effect known as network {\em densification}~\cite{LeCleFa07}. Specifically, the average degree in this case is given by
\begin{equation}
\overline{k(t)} \approx \frac{4T}{\sin{T\pi}}\frac{1}{1-\beta}\left[1-\frac{1}{t}+\frac{1}{\beta t}-\frac{1}{\beta t^{1-\beta}}\right]\xrightarrow[\beta\to1]{}\frac{4T}{\sin{T\pi}}\left[\ln{t}-1+\frac{1}{t}\right],
\end{equation}
where $\gamma=1+\frac{1}{\beta}$ is the exponent of the degree distribution as before. We see that the average degree grows logarithmically with the network size if $\gamma \to 2$. More generally, if $R_t=\delta r_t$ with $\delta \geq 1$, then we have
\begin{equation}
\overline{k(t)} \approx \frac{4T}{\sin{T\pi}}\frac{1}{1-\beta}\left[\frac{1}{\delta t^{1-\delta}}-\frac{1}{t\delta}+\frac{1}{(\beta+\delta-1)t}-\frac{1}{(\beta+\delta-1) t^{2-\delta-\beta}}\right],
\end{equation}
so that for large $t$ and $\gamma \rightarrow 2$, the average degree grows polynomially with the network size, $\overline{k(t)} \sim t^{\delta-1}\ln{t}$, if $\delta>1$. In this case the average shortest path distance and effective diameter do not increase but decrease with the network size, thus reproducing the shrinking diameter effect~\cite{LeCleFa07}, see Fig.~S9.

\begin{figure*}
\centerline{\includegraphics [width=3in]{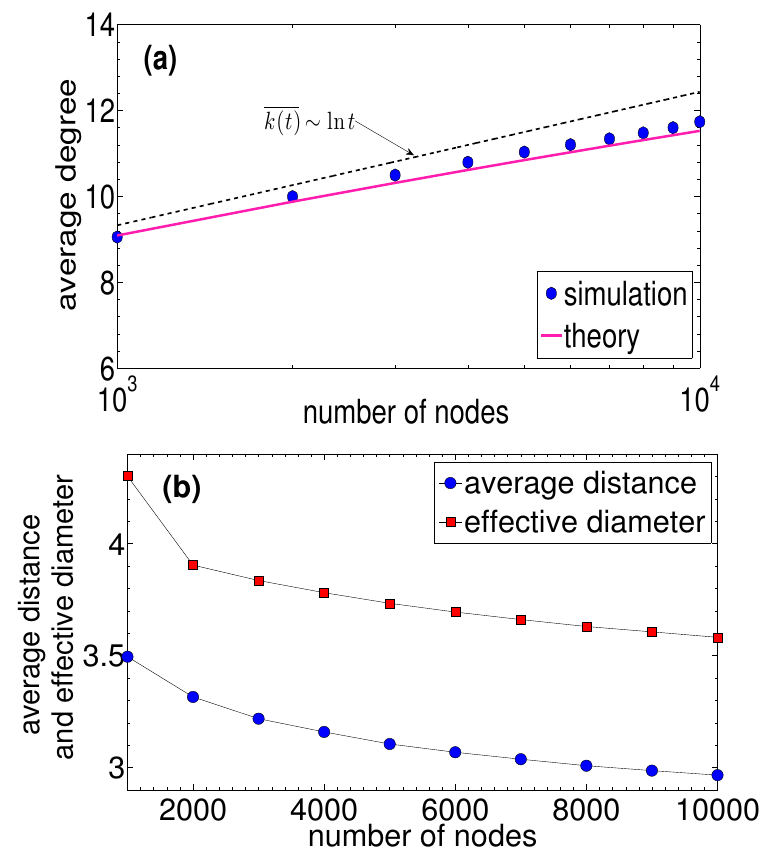}}
\caption{Densification effects. Plot~(a) shows the average degree $\overline{k(t)}$ as a function of size $t \in [10^3, 10^4]$ of networks grown according to $\textnormal{Model}_{2'}$, with $T=0.5, \gamma=2.1$, and $R_t=r_t$. Plot~(b) shows the average shortest path distance between nodes as a function of size $t \in [10^3, 10^4]$ of networks grown as in Plot~(a) but with $R_t=\delta r_t$, and $\delta=1.2$. The plot also shows the \emph{effective diameter} defined as the $90$th percentile of the shortest path distance distribution~\cite{LeCleFa07}.}
\end{figure*}

\section{Connection to the fitness model}
\label{sec:a2}

In this section we consider the popular fitness model~\cite{BiBa01a} and show that it can be also mapped to our geometric optimization framework.

The main motivation behind the fitness model is that in some real networks the popularity of a node does not depend only on its birth time,
but also on its ability (fitness) to compete for links.
Examples include the Web, where new sites may attract considerably more links than old ones, social networks
where new individuals may have more friends, and citation networks where new research papers may acquire a large number of citations
quickly.

To account for the different ability of nodes to compete for links in the fitness model~\cite{BiBa01a}, the following
attraction probability is introduced
\begin{equation}
\Pi(k_{\eta_s}(t))=m\frac{\eta_s \left(k_{\eta_s}(t)-m+A\right)}{\int_{1}^{t}\eta_i \left(\overline{k_{\eta_i}(t)}-m+A\right) di},
\label{eq:con_prob_fitness_1}
\end{equation}
which is a variant of Equation (\ref{eq:con_prob_pa_1}).
Equation (\ref{eq:con_prob_fitness_1}) says that the probability that an existing node $s$, $s < t$, attracts a link from a new node $t$
depends both on the node current degree $k_{\eta_s}(t)$ and on its fitness $\eta_s$. Fitness $\eta_s \in (0, \eta_{max}]$ is a parameter assigned
to each incoming node $s$, which remains unchanged in time and follows some distribution $\rho(\eta)$~\cite{BiBa01a}.
Given the fitness of each node, the attraction probability in Equation (\ref{eq:con_prob_fitness_1}) is conditioned on the
exact value of the degree of the node $k_{\eta_s}(t)$, and the unconditional probability can be obtained
after replacing $k_{\eta_s}(t)$ with its expected value $\overline{k_{\eta_s}(t)}$. We thus have
\begin{equation}
\Pi_{\textnormal{fitness}}(s,t)=m\frac{\eta_s \left(\overline{k_{\eta_s}(t)}-m+A\right)}{\int_{1}^{t} \eta_i \left(\overline{k_{\eta_i}(t)}-m+A\right)di}.
\label{eq:con_prob_fitness_2}
\end{equation}

Switching to our geometric optimization framework, to account for the fact that the popularity of different nodes can be
changing differently with time, we let nodes move with different speeds. That is, our model and its variants remain exactly the same,
with the only difference that every existing node $s$, $s < t$, now drifts away by increasing its radial coordinate using the
formula $r_s(t)=\beta(\eta_s) r_s +(1-\beta(\eta_s))r_t-\ln{\frac{\eta_s}{\eta_{max}}}$. Parameter $\beta(\eta_s)$ is some function of the fitness
of node $s$, $\eta_s$, and therefore its value can be different for different nodes.
We call this variant $\textnormal{Model}_3$.

Following exactly the same steps as in our earlier analysis, e.g., for $\textnormal{Model}_2$, we can see that
\begin{eqnarray}
\label{eq:con_prob_model_3}
\Pi_{\textnormal{Model}_3}(s,t) &=& m\frac{e^{-r_s(t)}}{\int_{1}^{t}e^{-r_i(t)}di}=m \frac{\eta_s\left(\frac{s}{t}\right)^{-\beta(\eta_s)}}{\int_{1}^{t} \eta_i\left(\frac{i}{t}\right)^{-\beta (\eta_i)}di},\\
\nonumber \overline{N(R_t)} &=& \frac{2T}{\sin{T\pi}} e^{-(r_t-R_t)}\frac{1}{\eta_{max}t} \int_{1}^{t}\eta_i \left(\frac{i}{t}\right)^{-\beta (\eta_i)}di,\\
R_t&=&r_t-\ln\left[\frac{2T}{\sin{T\pi}}\frac{\frac{1}{\eta_{max}t}\int_{1}^{t} \eta_i
\left(\frac{i}{t}\right)^{-\beta(\eta_i)}di}{m}\right] \quad\textnormal{for $\overline{N(R_t)}=m$.\label{eq:N_R_t_T_model_3}}
\end{eqnarray}
Parameter $T \in[0, 1)$ can be used again to tune clustering, and the limit $T \rightarrow 0$ is again well defined.

The integral $I(t)=\int_{1}^{t} \eta_i \left(\frac{i}{t}\right)^{-\beta (\eta_i)}di$
is in general a random variable that depends on the sequence of $\eta_i$'s, $i \in (1, t)$, and on the function $\beta(\eta)$.
As in~\cite{BiBa01a}, we compute the expected value of $I(t)$
\begin{equation}
\label{eq:I_t}
\overline{I(t)}=\int_{1}^{t}\int_{0}^{\eta_{max}}\eta \left(\frac{i}{t}\right)^{-\beta (\eta)}\rho(\eta)d{\eta}di \approx t C \quad\textnormal{for large}~t,
\end{equation}
where $C=\int_{0}^{\eta_{max}}\frac{\eta \rho(\eta)}{1-\beta (\eta)} d{\eta}$, and assume that $I(t) \approx \overline{I(t)}$.
We then get from Equation (\ref{eq:con_prob_model_3}) that
\begin{equation}
\label{eq:con_prob_model_3_approx}
\Pi_{\textnormal{Model}_3}(s,t) \approx \frac{m\eta_s}{t C}\left(\frac{s}{t}\right)^{-\beta(\eta_s)}.
\end{equation}
Using Equation (\ref{eq:con_prob_model_3_approx}) we compute the average degree of an existing node $s$ at time $t$, given its fitness $\eta_s$
\begin{eqnarray}
\nonumber \overline{k_{\eta_s}(t)} &=& m + \int_{s}^{t}\Pi_{\textnormal{Model}_3}(s,i) di \approx m+ \frac{m \eta_s}{\beta(\eta_s)C}\left[\left(\frac{s}{t}\right)^{-\beta(\eta_s)}-1\right]\label{eq:k_s_t_model_3}\\
&=& m+ A\left[\left(\frac{s}{t}\right)^{-\beta(\eta_s)}-1\right], \quad\textnormal{for $\beta(\eta_s)=\frac{m\eta_s}{AC}$.}
\end{eqnarray}
Observe that Equation (\ref{eq:k_s_t_model_3}) is similar to Equation (\ref{eq:k_s_t}) with the difference that the exponent is $\beta(\eta_s)$
instead of $\beta$, however, we again have $\overline{k_{\eta_s}(t)}\sim e^{-(r_s(t)-r_t)}$.
Using Equation (\ref{eq:k_s_t_model_3}) in (\ref{eq:con_prob_model_3}) we can see that
\begin{equation}
\Pi_{\textnormal{Model}_3}(s,t) = \Pi_{\textnormal{fitness}}(s,t).
\end{equation}
This means that for $m ,A, \rho(n)$ fixed, and $\beta(\eta)=\frac{m\eta}{AC}$, the probability that node $s$ attracts
a link from a new node $t$ is the same between $\textnormal{Model}_3$ and the fitness model, which in turn means that the resulting degree distribution is the same.
The degree distribution $P(k)$ is a weighted sum of different power laws, which can be computed following the approach in~\cite{BiBa01a}
\begin{equation}
\label{eq:p_k_fitness}
P(k)=\int_{0}^{\eta_{max}}d{\eta}\rho(\eta)\frac{C}{m\eta}\left(\frac{A}{k-m+A}\right)^{\frac{AC}{m\eta}+1}.
\end{equation}
Note that the attraction probability we consider in Equation (\ref{eq:con_prob_fitness_1}) is more general than the one used in~\cite{BiBa01a} and degenerates to it when $A=m$. In this case, we see that $\overline{k_{\eta_s}(t)}=m\left(\frac{s}{t}\right)^{-\beta(\eta_s)}$, $\beta(\eta_s)=\frac{\eta_s}{C}$, and $P(k)=\int_{0}^{\eta_{max}}d{\eta}\rho(\eta)\frac{C}{m\eta}\left(\frac{m}{k}\right)^{\frac{C}{\eta}+1}$, as in~\cite{BiBa01a}.

We conclude this section with some additional observations. As in~\cite{BiBa01a}, we conclude from Equation (\ref{eq:k_s_t_model_3}) that the exponent $\beta(\eta_s)$ is bounded, i.e., $0 < \beta(\eta_s) < 1~\forall s$, since a node always increases the number of links attached to it with time, $\beta(\eta_s) > 0$, and $\overline{k_{\eta_s}(t)}$ cannot increase faster than $t$, $\beta(\eta_s) < 1$. This means that $r_s(t)=\beta(\eta_s) r_s +(1-\beta(\eta_s))r_t-\ln{\frac{\eta_s}{\eta_{max}}} > 0,~\forall s$, as needed. Further, with $\beta(\eta)=\frac{m\eta}{AC}$ and
$A=(\gamma-2)m$, the value of $C$ is computed by the following Equation
\begin{equation}
1=(\gamma-2)\int_{0}^{\eta_{max}} \frac{\rho(\eta)}{\frac{(\gamma-2)C}{\eta}-1}d{\eta}.
\end{equation}
Since $\beta(\eta)=\frac{\eta}{(\gamma-2) C}< 1$ the singularity in the above integral is never reached and we also see that
$\eta_{max} < (\gamma-2)C$. Finally, when $\rho(\eta)=\delta(\eta-\tilde{\eta})$, i.e., all fitness equal to some $\tilde{\eta}$, $C=\frac{\gamma-1}{\gamma-2}\tilde{\eta}$ and $\beta(\tilde{\eta})=\beta=\frac{1}{\gamma-1}$ as expected, since in this case $\Pi_{\textnormal{Model}_3}(s,t)=\Pi_{\textnormal{fitness}}(s,t)=\Pi_{\textnormal{PA}}(s,t)$,
i.e., the degree distribution is the same, as if the network was growing according to standard preferential attachment
with power-law degree distribution exponent $\gamma$.

\section{Extensions for any curvature and temperature}
\label{sec:zeta_extension}

The general formula that gives the hyperbolic distance between two points $(r_s, \theta_s)$ and $(r_t, \theta_t)$ for any value of hyperbolic space curvature $K=-\zeta^2$, $\zeta > 0$ is~\cite{Bonahon09-book}
\begin{equation}
\label{eq:x_st_zeta}
x_{st}=\frac{1}{\zeta}\textnormal{arccosh}(\cosh{\zeta r_s}\cosh{\zeta r_t}-\sinh{\zeta r_s} \sinh {\zeta r_t} \cos{\theta_{st}}) \approx r_s+r_t+\frac{2}{\zeta}\ln(\theta_{st}/2).
\end{equation}
The popularity$\times$similarity model with $T \in [0,1)$ can be extended to any $\zeta < \infty$ with the following two simple
modifications: (i) the initial radial coordinate of each new node $t \geq 1$ is $r_t=\frac{2}{\zeta}\ln{t}$ (instead of $\ln{t}$); and (ii)
given the hyperbolic distance $x_{st}$ between new node $t$ and existing node $s$, node $t$ connects to $s$ with probability
$p(x_{st})=1/[1+e^{\zeta(x_{st}-R_t)/(2T)}]$ (instead of $p(x_{st})=1/[1+e^{(x_{st}-R_t)/T}]$). Redoing the analysis for $\textnormal{Model}_{2}$ (or
$\textnormal{Model}_{2'})$ it is easy to check
that exactly the same results hold, and that $R_t$ is now given by the more
general formula
\begin{equation}
R_t=r_t-\frac{2}{\zeta}\ln\left[\frac{2T}{\sin{T\pi}}\frac{\left(1-e^{-\frac{\zeta}{2}(1-\beta)r_t}\right)}{m (1-\beta)}\right],
\label{eq:R_t_T_zeta}
\end{equation}
with the limit $T \rightarrow 0$ again well defined. The expected degree of node $s$ at time $t$ in this case is $\overline{k_s(t)} \sim e^{-\frac{\zeta}{2}(r_s(t)-r_t)}$.

The extension for any $T > 1$ is a bit more involved, but we need it for the next section
where we consider interesting high-temperature limits. The point $T=1$ is a phase transition and
for $T \geq 1$ the approximation in Equation (\ref{eq:P_s_t}) giving $P(s,t)$ no longer
holds. In particular, after performing the change of variables $\chi_{st}=X(s,t)\frac{\theta_{st}}{2}$ as in Equation (\ref{eq:c_v_1}), we see that
the corresponding integral (Equation (\ref{eq:c_v_2})) diverges, and we explicitly have to cut off the integration at the maximum value $X (s, t)
\frac{\pi}{2}$. This yields for $T > 1$
\begin{equation}
\label{eq:P_s_t_T_g_1}
P(s,t) \approx \left(\frac{2}{\pi}\right)^{\frac{1}{T}}\frac{T}{t(T-1)}\frac{1}{[X(s,t)]^\frac{1}{T}},
\end{equation}
with $X(s,t)=e^{\frac{\zeta}{2}(r_s(t)+r_t-R_t)}$, $\zeta > 0$. In this high-temperature regime, the model has the same attraction probability and degree distribution as in the low-temperature regime $T < 1$
if the initial radial coordinate of each new node $t \geq 1$ is
$r_t=\frac{2T}{\zeta}\ln{t}$ instead of $r_t=\frac{2}{\zeta}\ln{t}$, yielding
\begin{equation}
\label{eq:R_t_T_g_1}
R_t=r_t-\frac{2T}{\zeta} \ln\left[\left(\frac{2}{\pi}\right)^{\frac{1}{T}}\frac{T}{T-1}\frac{1-e^{-\frac{\zeta}{2T}(1-\beta)r_t}}{m(1-\beta)}\right].
\end{equation}
We can now allow $\zeta \rightarrow \infty$ if at the same time $T \sim \zeta \rightarrow \infty$.
The main difference compared to $T < 1$ is that clustering is asymptotically zero for any $T>1$.
We have confirmed this effect and all the expressions in this section in simulations.

\section{Connections to preferential attachment, growing random graphs, and growing random geometric graphs}
\label{sec:a4}

In this section we show that standard PA with asymptotically zero clustering~\cite{BoRi02}, growing random graphs~\cite{SoRa51} and growing random geometric graphs~\cite{Penrose03-book}, can all be seen as limiting degenerate cases of popularity$\times$similarity optimization.

To see the connection to standard PA, we need to consider the general formula that gives the hyperbolic distance $x_{st}$ between
two points $(r_s, \theta_s)$ and $(r_t, \theta_t)$ for any value of hyperbolic space curvature $K=-\zeta^2$, $\zeta > 0$,
given by Equation (\ref{eq:x_st_zeta}). By letting curvature go to minus infinity, $\zeta \rightarrow \infty$, we transform the hyperbolic space to a tree~\cite{Bonahon09-book}, and kill the $\theta$-dependent term in the expression for $x_{st}$~(\ref{eq:x_st_zeta}), i.e., the term abstracting the similarity distance.
That is, the hyperbolic distance between nodes depends only on their popularity, $x_{st}=r_s+r_t$, as in PA. We can now set $T\sim\zeta$, e.g., $T=\frac{\zeta}{2}$ without loss of generality. This setting yields $r_t=\frac{2T}{\zeta}\ln{t}=\ln{t}$, and Equation (\ref{eq:R_t_T_g_1}) becomes
\begin{equation}
R_t=r_t-\ln\left[\frac{1-e^{-(1-\beta)r_t}}{m (1-\beta)}\right].
\label{eq:R_t_T_limit_pa}
\end{equation}
Further, from Equation (\ref{eq:P_s_t_T_g_1}), the connection probability is now
\begin{equation}
P(s,t)\approx \frac{1}{t}e^{-(r_s(t)+r_t-R_t)},
\label{eq:P_s_t_limit_pa}
\end{equation}
so that the probability that node $s$ attracts a link from node $t$ is again
\begin{equation}
m\frac{P(s,t)}{P(t)}=m\frac{e^{-r_s(t)}}{\int_{1}^{t}e^{-r_i(t)}di}=m\frac{\left(\frac{s}{t}\right)^{-\beta}}{\int_{1}^{t}\left(\frac{i}{t}\right)^{-\beta}di}=\Pi_{PA}(s,t).
\label{eq:con_prob_model_2_limit_pa}
\end{equation}
As before, this means that the degree distribution is a power law with exponent $\gamma=1+\frac{1}{\beta}$, but clustering is zero since $T\to\infty$.
Figure~S10 shows simulation results validating our analysis.

\begin{figure*}
\centerline{\includegraphics [width=3in]{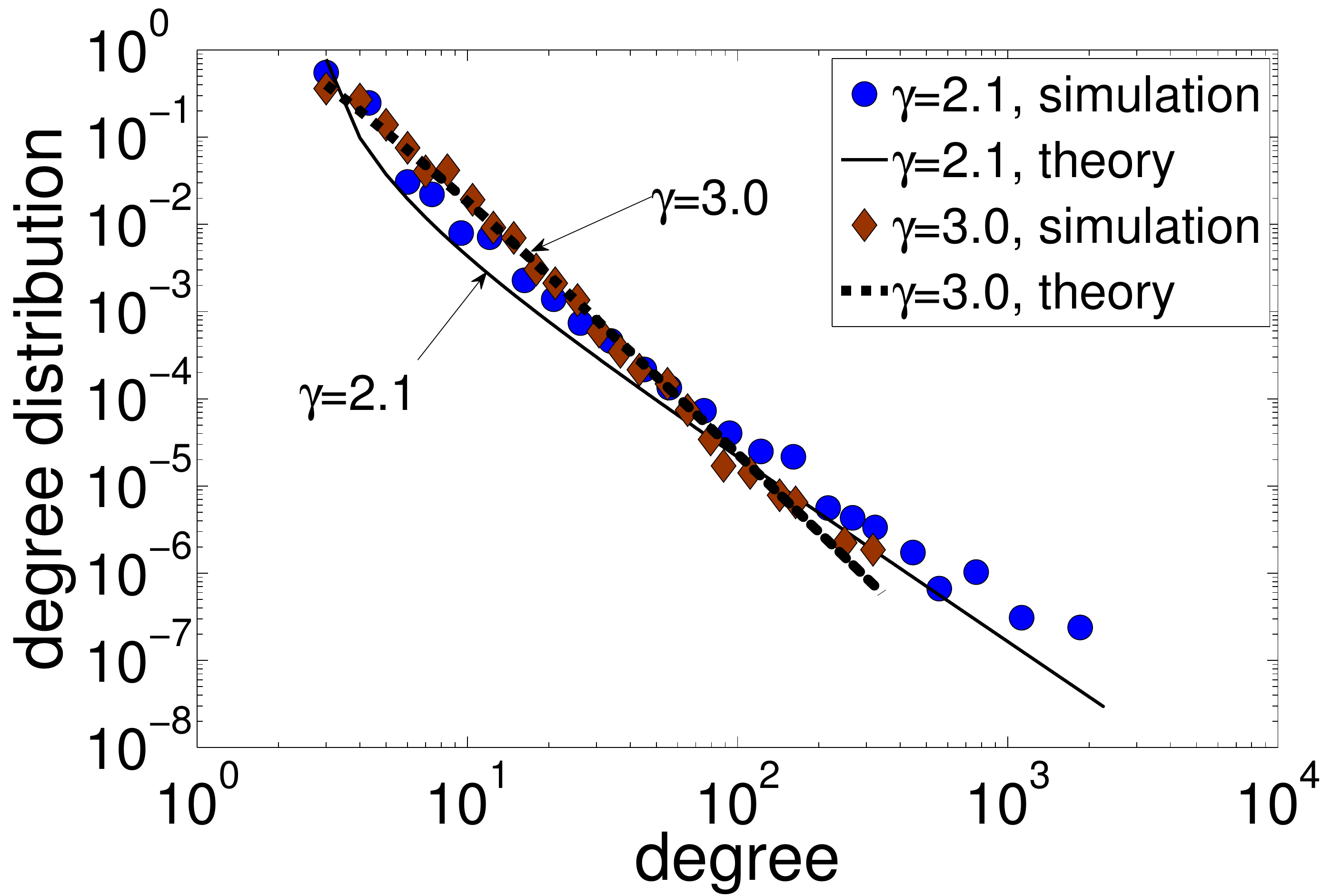}}
\caption{Distribution $P(k)$ of node degrees in networks growing according to the standard PA limit with $T=\frac{\zeta}{2} \rightarrow \infty$, $m=3$, and $\gamma=2.1, 3.0$.
The theoretical predictions are given by Equation (\ref{eq:P_k_theory}). For $\gamma=2.1, 3.0$ the average clustering in the simulated networks is $\bar{c}= 0.068, 0.004$.}
\end{figure*}

If we now let $\beta \rightarrow 0$, then $\gamma \rightarrow \infty$, and the generated networks
degenerate to growing random graphs. Indeed, if $\beta=0$, $r_s(t)=r_t$, $\forall s,t$,
i.e., all node pairs have the same popularity as they all lie on the circle of the maximum
radius $r_t$, expanding with time. It is easy to check that the attraction probability is now
$m\frac{P(s,t)}{P(t)}=\frac{m}{t-1} \approx \frac{m}{t}$, $\forall s, t$.
This probability is similar to the connection probability in classical random graphs ${\cal G}_{N,p}$~\cite{SoRa51},
where each $N(N-1)/2$ pair of $N$ nodes is connected with the same probability $p\approx\bar{k}/N$. The difference is that our
graphs are growing, which affects their properties including the degree distribution.
The degree distribution in these growing graphs is exponential~\cite{DorMen-book03}, versus the Poissonian distribution
in classical random graphs.

The limit $\beta \rightarrow 0$ ($\gamma \rightarrow \infty$) also exists at low temperatures $T \in [0, 1)$ with finite clustering
controlled by $T$. In this case, we can check that the attraction probability is still $\frac{m}{t}$, $\forall s, t$, as all nodes are equally
popular, but clustering is not zero, as similarity (the angular distance between nodes) matters.
When $T=0$ we have the strongest clustering, and the generated networks degenerate to growing random geometric graphs on the circle. Indeed, we see from Equation (\ref{eq:R_t_prob}) that since any two nodes $s, t$ have the same radial coordinate $r_t$, they are connected only if the distance between them on the circle is less than a constant that depends on $t$, i.e., $t$ connects to $s$ only if $\theta_{st} \leq 2 e^{-\left(2r_t-R_t\right)}=\frac{m\pi}{t-1} \approx \frac{m\pi}{t}$.

In equilibrium geometric networks~\cite{KrPa10}, the connections to PA, growing random graphs, and growing random geometric graphs,
are, respectively, the connections to the soft configuration model (random graphs with a given expected degree distribution), classical random graphs, and random geometric graphs.

\section{Extension with internal links}
\label{sec:internal_extension}

While in some real networks, e.g., citation networks, new connections appear only from new to old nodes, in some other networks, new links may connect pairs of old, previously disconnected nodes. These links are called \emph{internal,} versus \emph{external\/} links of the previous type. Examples of networks with internal links include the Internet, were existing disconnected ASs may decide to connect at some point, and social networks were existing disconnected individuals may become friends or collaborators. Our geometric optimization framework can be easily extended to account for internal links as we show below.

At each time $t$, in addition to the $m$ external links introduced by new node $t$ (e.g., using $\textnormal{Model}_{2}$), $L$ internal connections are also created between existing disconnected pairs of nodes. Specifically, a random pair of existing nodes $i, j < t$  is selected, and then connected (given that it is disconnected) with probability $p(x_{ij})=1/[1+e^{(x_{ij}-R_t)/T}]$. The step is repeated until $L$ internal links are created. This procedure is exactly the same as the procedure by which a new node $t$ connects to existing nodes in $\textnormal{Model}_{2}$. The average degree is now $\bar{k}=2(m+L)$.

Following exactly the same procedure as in Section~\ref{sec:degree_distribution}, and considering any value of hyperbolic space curvature $K=-\zeta^2$, $\zeta > 0$ (Section~\ref{sec:zeta_extension}), the probability that existing nodes $i, j$ are selected and connected at time $t$ is
\begin{equation}
P(i,j, t) \approx \frac{2T}{t^2\sin{T\pi}}\frac{1}{X(i,j,t)}, \quad\textnormal{where~} X(i,j,t)=e^{\frac{\zeta}{2}(r_i(t)+r_j(t)-R_t)}.
\label{eq:P_i_j_t}
\end{equation}
The probability that a pair of existing nodes gets connected at time $t$ is $\frac{1}{2}\int_{1}^{t}\int_{1}^{t}P(i, j, t)didj$. Since $L$ internal links are introduced, the probability that pair $i, j$ attracts a link is
\begin{equation}
\Pi(i,j, t)=2L\frac{P(i,j,t)}{\int_{1}^{t}\int_{1}^{t}P(i, j, t)didj}.
\label{eq:attraction_prob_i_j}
\end{equation}
Therefore, the probability that node $s < t$ attracts an internal link at time $t$ is
\begin{equation}
\label{eq:con_prob_internal}
\Pi^{\textnormal{internal}}(s,t)=\int_{1}^{t}\Pi(s,i,t)di=2L\frac{e^{-\frac{\zeta}{2}r_s(t)}}{\int_{1}^{t}e^{-\frac{\zeta}{2}r_i(t)di}}=2L\frac{\left(\frac{s}{t}\right)^{-\beta}}{\int_{1}^{t}\left(\frac{i}{t}\right)^{-\beta}di},
\end{equation}
which is similar to the probability that node $s$ attracts an external link, with the only difference that here we have the prefactor $2L$ instead of $m$, see Equation (\ref{eq:con_prob_model_2}). Thus, the total probability that node $s$ attracts a link at time $t$ is the probability that the node attracts an external or an internal link
\begin{equation}
\label{eq:con_prob_total}
\Pi^{\textnormal{total}}(s,t)=(m+2L)\frac{\left(\frac{s}{t}\right)^{-\beta}}{\int_{1}^{t}\left(\frac{i}{t}\right)^{-\beta}di}=(\bar{k}-m)\frac{\left(\frac{s}{t}\right)^{-\beta}}{\int_{1}^{t}\left(\frac{i}{t}\right)^{-\beta}di}.
\end{equation}
The average degree of node $s$ by time $t$ is now given by
\begin{equation}
\label{eq:k_s_t_with_internal}
\overline{k_s(t)}=m+A'\left[\left(\frac{s}{t}\right)^{-\beta}-1\right],
\end{equation}
where $A'=(\bar{k}-m)(\gamma-2)$. Equation (\ref{eq:k_s_t_with_internal}) is similar to Equation (\ref{eq:k_s_t}), and is identical to it if $L=0$ (i.e., if $\bar{k}=2m$). From Equations (\ref{eq:con_prob_total}) and (\ref{eq:k_s_t_with_internal}) we see that a node of degree $k$ attracts a new
link at time $t$ with probability
\begin{equation}
\label{eq:con_prob_pa_with_internal}
\Pi(k)=(\bar{k}-m)\frac{k-m+A'}{(\bar{k}-m+A')t}.
\end{equation}

The link attraction probabilities in Equations (\ref{eq:con_prob_pa_with_internal}) and (\ref{eq:pi_k_theory}) are identical if $L=0$. If $L=\frac{\bar{k}-2m}{2} > 0$, Equation (\ref{eq:con_prob_pa_with_internal}) gives approximately the same probability as Equation (\ref{eq:pi_k_theory}) for sufficiently large $k$, i.e., for $k \geq \frac{\bar{k}}{2}$, and the absolute difference between the two probabilities is $\frac{L}{t}$. This observation implies that for a target $\bar{k}$ and $\beta=\frac{1}{\gamma-1}$ the degree distributions in both cases are nearly identical, and indistinguishable from the degree distribution in networks growing according to standard PA. However, while internal links do not affect the degree distribution, they can affect other topological characteristics, e.g., they can decrease the average distance in the network. We study topological characteristics of networks growing according to popularity$\times$similarity optimization with internal links in the next section.

We conclude this section with some additional notes. First, from the analysis above we see that, similar to external links, PA appears as an emergent effect in the internal link attraction probability as well since a node attracts an internal link with probability which is also proportional to its current degree. Second, temperature $T$ has the same effect on internal connections as on external connections, i.e., smaller values of $T$ increase the probability that hyperbolically close disconnected node pairs get connected, which increases clustering. Finally, the model extension with internal links can be combined with the fitness model extension, described in Section~\ref{sec:a2}, as the former does not depend on whether nodes are moving with the same speeds or not. In this combination Equation (\ref{eq:con_prob_internal}) becomes $\Pi^{\textnormal{internal}}(s,t)=2L\,e^{-\frac{\zeta}{2}r_s(t)}/\int_{1}^{t}e^{-\frac{\zeta}{2}r_i(t)di}=2L\eta_s\left(\frac{s}{t}\right)^{-\beta(\eta_s)}/\int_{1}^{t}\eta_i\left(\frac{i}{t}\right)^{-\beta (\eta_i)}di$, which is similar to Equation (\ref{eq:con_prob_model_3}), and a straightforward analysis as above can be applied.

\section{Properties of real-world versus modeled networks}
\label{sec:dk_compare}

In this section we compare several important properties of the real-world networks considered in Section~\ref{sec:real_nets} to the properties of modeled networks growing according to popularity$\times$similarity optimization. Specifically, we consider the following properties:
\begin{enumerate}
\item[(a)] degree distribution $P(k)$;
\item[(b)] average clustering $\bar{c}(k)$ of $k$-degree nodes;
\item[(c)] average degree of neighbors $\bar{k}_{nn}(k)$ of $k$-degree nodes;
\item[(d)] distance distribution $d(l)$, i.e., the distribution of hop lengths $l$ of shortest paths between nodes in the network, or the probability that a random pair of nodes are at the distance of $l$ hops from each other;
\item[(e)] average node betweenness $\bar{B}(k)$ of $k$-degree nodes, which is the average number of shortest paths passing through
a $k$-degree node, normalized by the maximum possible number of such paths.
\end{enumerate}
Property~(c) captures \emph{degree correlations} in the network. If $\bar{k}_{nn}(k)$ is an increasing function, then high (low) degree nodes connect, on average, to nodes of high (low) degree, and the network is called \emph{assortative}. Otherwise, nodes of high degree tend to connect to nodes of low degree, and the network is called \emph{disassortative}. Technological and biological networks are usually disassortative, while social networks are usually assortative~\cite{Dorogovtsev10-book,Newman10-book}. Properties~(a-c) are {\em local\/} statistics reflecting properties of individual nodes and their one-hop neighborhoods, as opposed to {\em global\/} properties~(d-e) which depend on large-scale organization of the network.

\subsection{Internet}
We take the Archipelago AS Internet topology of June 2009 from Section~\ref{sec:internet_data}, and compute properties (a)$...$(e) from above.
The network consists of $t=23748$ nodes, and has $\gamma=2.1$, $\bar{k}\approx 5$, $\bar{c}=0.61$.  Then we grow a network according to
the popularity$\times$similarity model ($\textnormal{Model}_{2'}$) up to the same number of nodes as in the real AS Internet, and with the same $\gamma$, $\bar{k}$ and $\bar{c}$. We compute the same properties in the resulting network, and compare them to those of the real Internet. The results are shown in Fig.~S11, where we observe a good match between the properties of the modeled network and real Internet. This match is even better if we allow for internal connections as described in Section~\ref{sec:internal_extension}. In this case, each new node connects on average to $m=1.5$ existing nodes, and at each time $L=1$ existing disconnected pairs of nodes are connected so that $\bar{k}=2(m+L)=5$. With no internal links, $L=0$ and $m=\bar{k}/2=2.5$.

\begin{figure*}
\centerline{\includegraphics [width=7.2in]{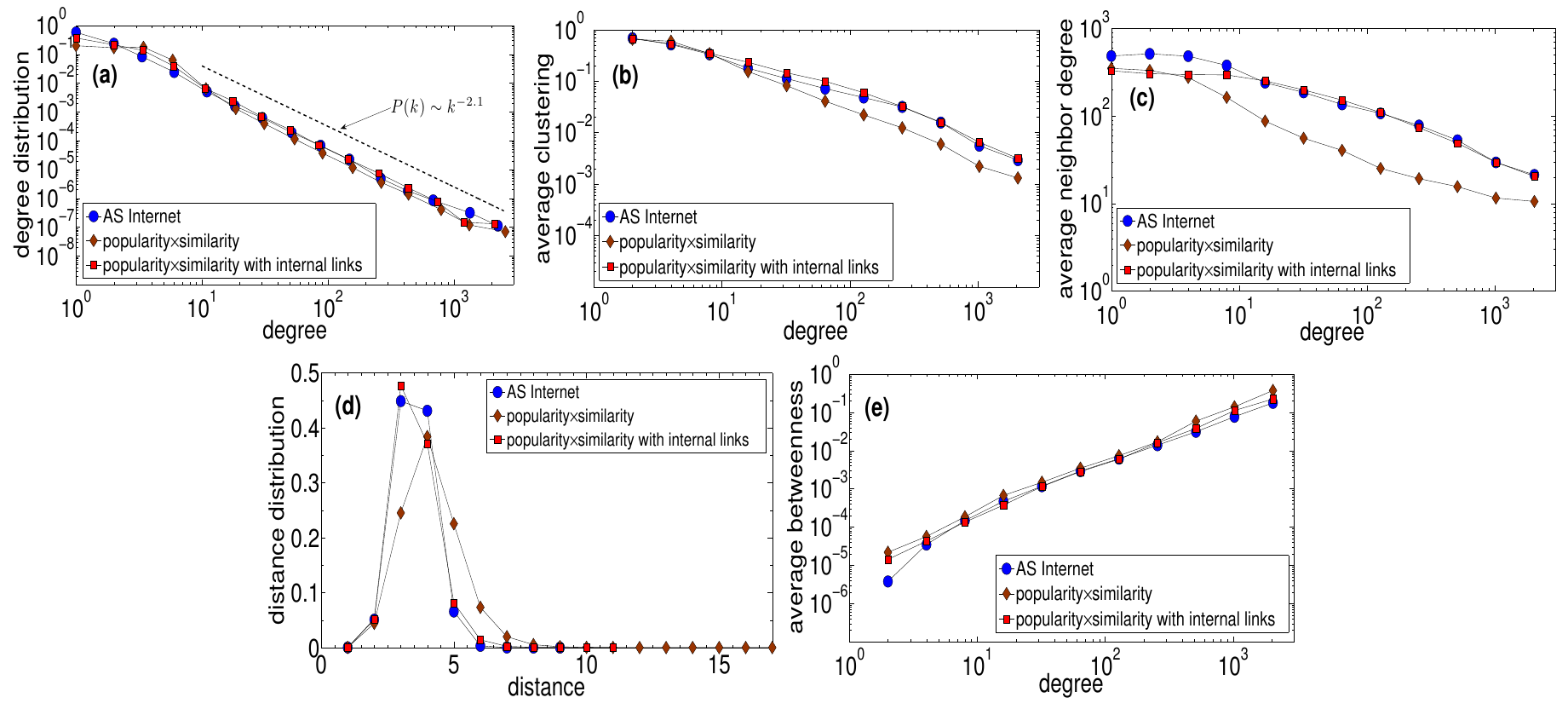}}
\caption{Properties of the AS Internet vs.\ networks
grown according to popularity$\times$similarity optimization. The plots show: (a) the degree distribution $P(k)$;
(b) the average clustering $\bar{c}(k)$ of $k$-degree nodes; (c) the average neighbor degree $\bar{k}_{nn}(k)$ of
$k$-degree nodes; (d) the distance distribution $d(l)$; and (e) the average node betweenness $\bar{B}(k)$ of $k$-degree
nodes.}
\end{figure*}

\subsection{{\it E.coli\/} metabolic network}

Here we consider the entire network of metabolites from Section~\ref{sec:metabolic_data}, and compute properties (a)$...$(e) for it. Recall that the network consists of $t=1010$ nodes, and has $\gamma=2.5$, $\bar{k}=6.5$, $\bar{c}=0.48$. We grow a network according to the popularity$\times$similarity model ($\textnormal{Model}_{2'}$) up to the same number of nodes as in the metabolic network, and with the same $\gamma$, $\bar{k}$, and $\bar{c}$. We use $m=\bar{k}/2=3.25$. We compute the same network properties in the resulting network, and compare them to those of the real metabolic network. The results are shown in Fig.~S12, where we observe a remarkable match across all five properties.

\begin{figure*}
\centerline{\includegraphics [width=7.2in]{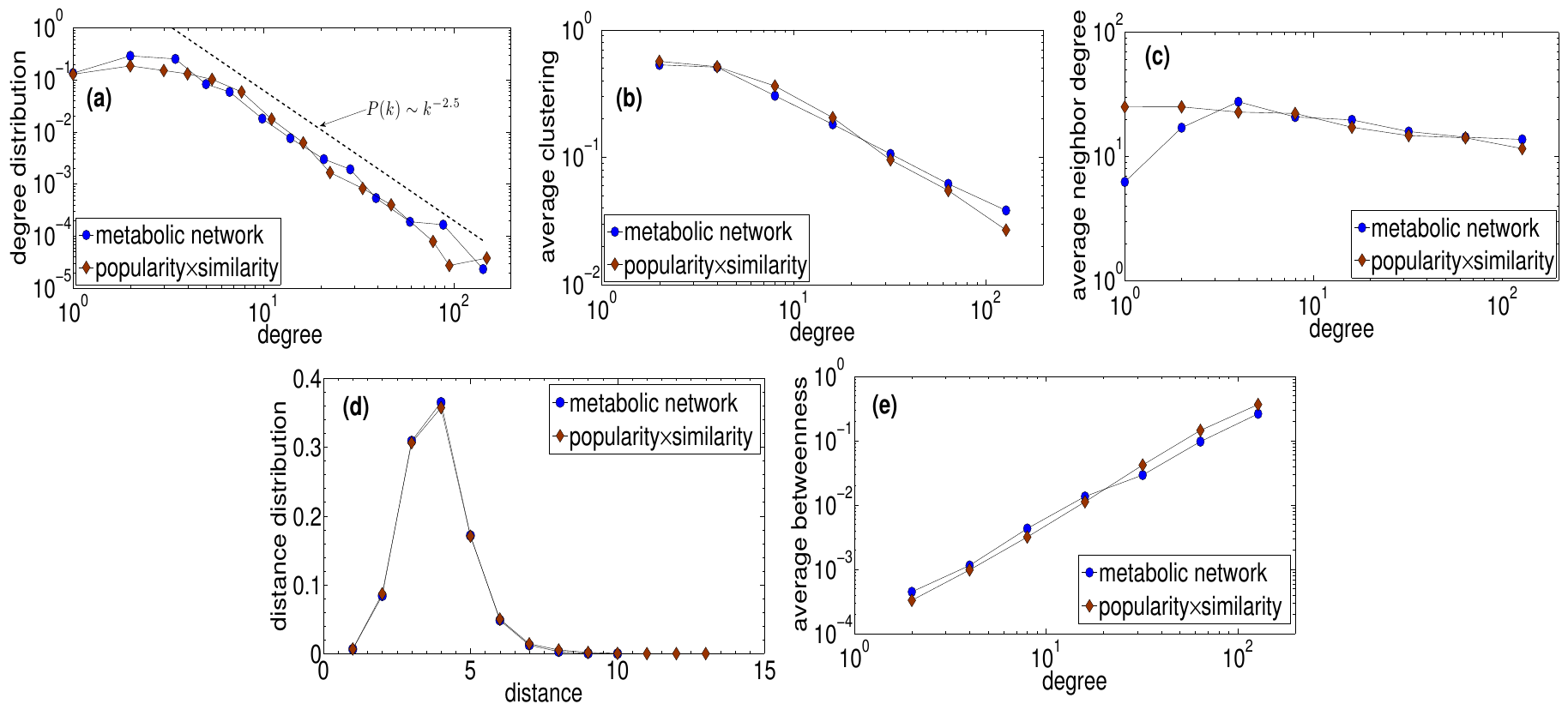}}
\caption{Properties of the {\it E.coli\/} metabolic network vs.\ networks
grown according to popularity$\times$similarity optimization. The plots show: (a) the degree distribution $P(k)$;
(b) the average clustering $\bar{c}(k)$ of $k$-degree nodes; (c) the average neighbor degree $\bar{k}_{nn}(k)$ of
$k$-degree nodes; (d) the distance distribution $d(l)$; and (e) the average node betweenness $\bar{B}(k)$ of $k$-degree
nodes.}
\end{figure*}

\subsection{PGP web of trust}

\begin{figure*}
\centerline{\includegraphics [width=7.2in]{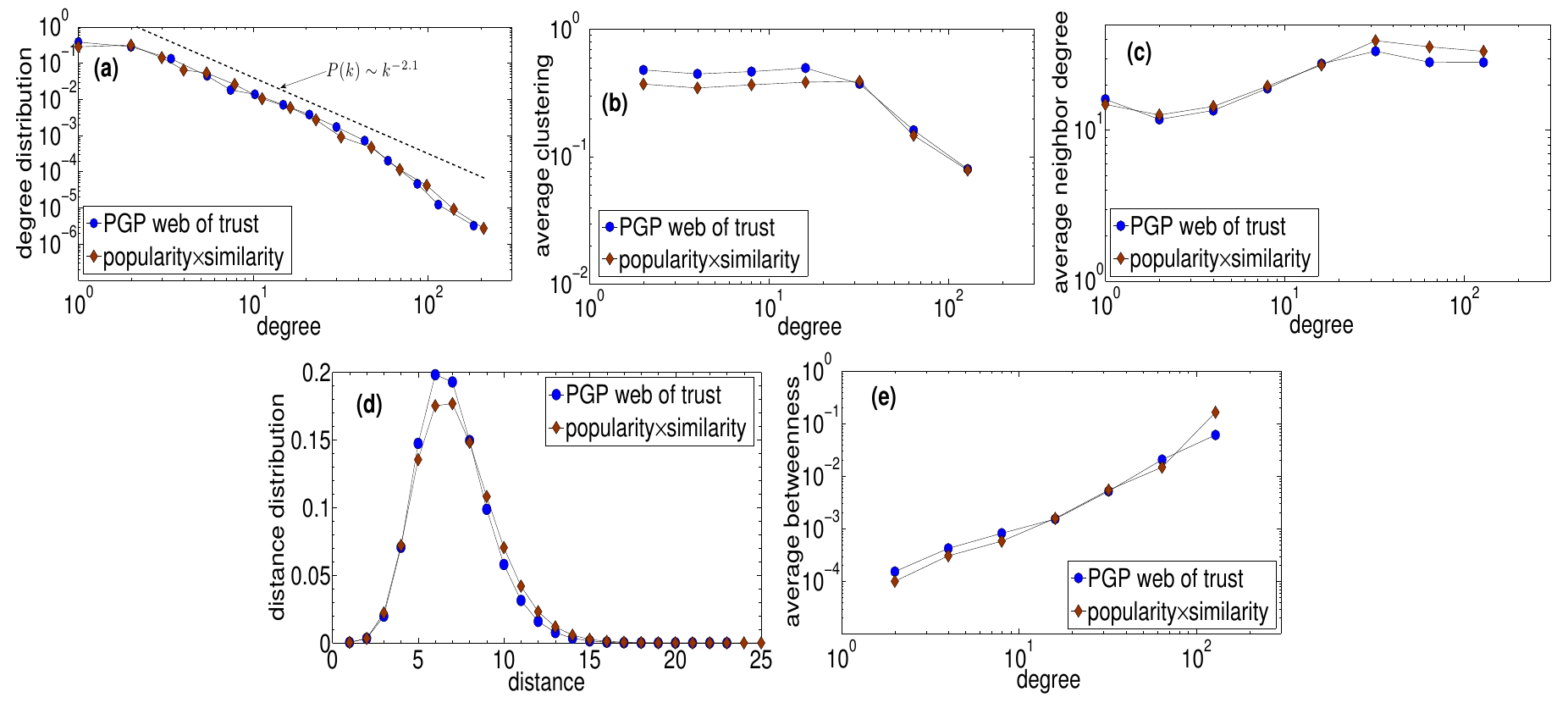}}
\caption{Properties of the PGP web of trust vs.\ networks
grown according to popularity$\times$similarity optimization. The plots show: (a) the degree distribution $P(k)$;
(b) the average clustering $\bar{c}(k)$ of $k$-degree nodes; (c) the average neighbor degree $\bar{k}_{nn}(k)$ of
$k$-degree nodes; (d) the distance distribution $d(l)$; and (e) the average node betweenness $\bar{B}(k)$ of $k$-degree
nodes.}
\end{figure*}

We now take the PGP web of trust snapshot of April 2003 from Section~\ref{sec:pgp_data}, and compute its properties (a)$...$(e). The network consists of $t=14367$ nodes, and has $\bar{k}=5.3$, $\bar{c}=0.47$. Its degree distribution is shown in Fig.~S13(a), where we observe deviations from a clean power law.

This observation motivates us to grow a modeled network using the fitness model extension in Section~\ref{sec:a2}, i.e., $\textnormal{Model}_3$, which can model non-power-law degree distributions. Recall that in $\textnormal{Model}_3$, nodes $s$, $s < t$, move with different speeds, increasing their radial coordinate according to $r_s(t)=\beta(\eta_s) r_s +(1-\beta(\eta_s))r_t-\ln{\frac{\eta_s}{\eta_{max}}}$, where
$\beta(\eta_s) \sim \eta_s$, and $\eta_s$ is the fitness of $s$. To grow a network according to this model,
we need to know $\beta(\eta_s),~\forall s \leq t$. Given that $\frac{\beta(\eta_s)}{\beta(\eta_{max})}=\frac{\eta_s}{\eta_{max}}$, we can find $\beta(\eta_s)$ by solving
\begin{equation}
\label{eq:beta_sol}
r_s(t)=\beta(\eta_s) r_s +(1-\beta(\eta_s))r_t-\ln{\frac{\beta(\eta_s)}{\beta(\eta_{max})}},
\end{equation}
since we know $r_t=\ln t$, have $r_s(t)$ inferred in Section~\ref{sec:mapping}, and can infer $r_s$ as follows. We assume that nodes with smaller current radial coordinates were born earlier, and sort them in the increasing order, thus creating a sequence of current inferred radial coordinates $r_1(t), r_2(t),...,r_t(t)$ for nodes born at times $s=1, 2,...,t$. Nodes for which the current radial coordinate is the same, are assumed to have appeared at the same time. Using $r_s=\ln{s}$, and setting $\beta(\eta_{max})=1$, we have all the ingredients to solve Equation~(\ref{eq:beta_sol}) for $\beta(\eta_s)$ for every node $s=1, 2,...,t$.

Another peculiarity of the PGP network, compared to the networks considered earlier, is a deviation of the distribution of the inferred angular distances between nodes from the uniform distribution: see Fig.~S14 showing these distributions for all the considered real networks. In the PGP network, nodes with small radial coordinates are, on average, at smaller angular distances than what the uniform distribution suggests. Therefore in growing the modeled PGP network, we use the inferred angular coordinate $\theta_s$ for every node $s=1, 2,...,t$, even though our analysis in Section~\ref{sec:a2} assumes a uniform angular distance distribution.

\begin{figure*}
\centerline{\includegraphics [width=7.2in]{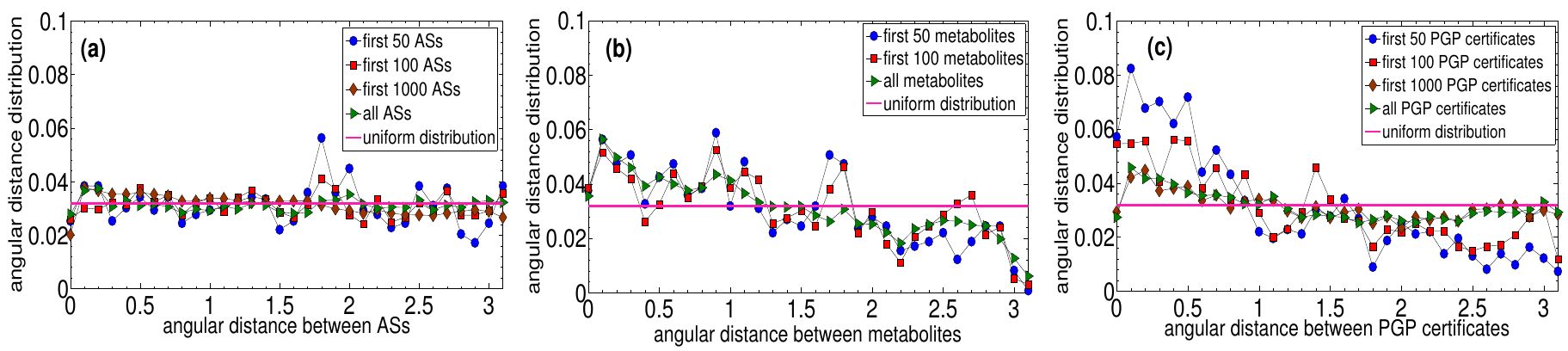}}
\caption{Distribution of the inferred angular distances (in radians) in the AS Internet (plot (a)), {\it E.coli\/} metabolic network (plot (b)),
and PGP web of trust (plot (c)). In each case, we first sort the nodes in the increasing order of their inferred radial coordinates, so that the first
$i$ nodes are the nodes with the $i$ smallest radial coordinates, and then compute the distribution of the angular distances for these first $i$ nodes, using
bins of size $\delta=0.1$. We vary $i$ from small values up to the total number of nodes in the network. The straight line in each plot is the uniform distribution $p(\theta)=\delta/\pi$.}
\end{figure*}

Figure~S13 juxtaposes properties (a)$...$(e) of a network grown according to $\textnormal{Model}_3$ up to $t=14367$ nodes using the inferred $\beta(\eta_s)$'s and $\theta_s$'s, temperature $T=0.2$, $m=1$, and $L=1.65$ ($\bar{k}=2(m+L)=5.3$), against the corresponding properties of the real PGP snapshot. As with the AS Internet and {\it E.coli\/} metabolic network, we also observe a good match between the modeled and real PGP web of trust across all these properties.\\

To summarize this section, synthetic networks growing according to popularity$\times$similarity optimization reproduce several important structural characteristics of real technological, biological, and social networks. Remarkably, this optimization approach can capture the properties of both disassortative (Figs.~S11(c), S12(c)) and assortative (Fig.~S13(c)) networks, as well as networks with degree distributions deviating from clean power laws (Fig.~S13(a) vs.\ Figs.~S11(a), S12(a)).

\section{Related work}
\label{sec:related_work}

\subsection{Optimization}

The work that comes perhaps closest to our approach is by D'Souza et al.~\cite{SoBo07,BeBo05}. In this work the authors show that PA can emerge in a tradeoff optimization framework requiring only local information. The framework is motivated by how connections in the Internet may take place. Specifically, the motivation is that a new AS may want to establish connections that would minimize the startup costs, while still providing good performance to its users. In the model, a new node is placed on the unit interval where distances abstract the connection fibre costs, and then connects to an existing node minimizing a balance between these costs and the shortest path hop-lengths to the core in the network, the latter abstracting performance in terms of the average delay from the new node to the rest of the network. The authors then focus only on the degree distribution in the graphs produced by this model, showing that with a specific fit of parameters, it matches well the degree distribution of the Internet extracted from the WHOIS data. The basic model studied in this work generates trees, since each incoming node connects to $m=1$ existing nodes, but the authors suggest at the end that for $m>1$ the model may lead to some non-zero clustering.

\subsection{PA$+$similarity information}

The fact that similarity between nodes affects the linking probability in networks has been observed, studied, and modeled extensively in the literature \cite{Redner98,McPh01,WatDoNew02,menczer02-pnas,menczer04-pnas,BoMaGo04-pnas,MuIt07,CraCo08,SiJe08,LiJi09,triangular_clustering}. Of particular interest are the works by Menczer~\cite{menczer02-pnas,menczer04-pnas} where he introduces a model for text corpora with linking probability that augments standard PA with document similarity measures. The latter can be the standard cosine similarity for a pair of documents, defined by the normalized count of words common to both documents. The author then shows that this model describes well the degree and similarity distributions in the DMOZ Web data and in a collection of articles published in PNAS. In~\cite{menczer02-pnas} he also shows that similarity information can help to improve Web navigation, an observation confirmed later in a more abstract context~\cite{SiJe08}, where similarity is modeled by distances on the unit interval. In~\cite{triangular_clustering} a modification of the model of~\cite{menczer04-pnas} is proposed where the linking probability is proportional to the product of the degrees of the documents and their cosine similarity. The authors then show that this model can describe the clustering coefficient in document networks better compared to~\cite{menczer04-pnas}. In~\cite{LiJi09} similarity attributes are modeled by vectors in an $n$-dimensional space. A new node first selects a certain group of existing nodes (community) based on similarity distances between the new and existing nodes. Within the community the attachment then follows standard PA. That is, this model also augments PA with similarity. The authors conclude by showing that the model generates graphs with power-law degree distributions and exponent $\gamma=3$, and some community structure. No real networks are considered.

\subsection{PA$+$spatial information}

A wider class of models augment PA not with similarity information {\it per se}, but with some spatial information~\cite{YoJeBa02,Dell'Amico2006,FeCo11}, see also Section~4.4 in~\cite{Barthelemy11}. In these models, nodes are located in some space, and the linking probability depends not only on node degrees as in standard PA, but also on distances between nodes in the space. If this linking probability decreases with the spatial distance fast enough, then such models generate graphs with strong clustering for a very simple reason: since close nodes have high probability of being connected, then the triangle inequality in the space leads to a large number of triangles in the network. Yet the mechanism responsible for power-law degree distributions in these models is the same PA.

\subsection{Hidden variables}

Yet wider and more general class of models, to which our approach actually belongs, are the network models with hidden variables~\cite{CaCaDeMu02,BoPa03}. In these models, some hidden variables are first assigned to nodes, and the linking probability between a pair of nodes is then a function of the values of their hidden variables. For example, in~\cite{CaCaDeMu02} the authors show that a combination of exponentially distributed hidden variables and step-function connection probability leads to power-law degree distributions and strong clustering in modeled networks, while in~\cite{BoPa03} it is shown that PA itself can be casted as a hidden variable model, where one of the hidden variables is the node birth time.

\subsection{Clustering}

A variety of other mechanisms have been proposed to fix the zero-clustering problem with PA. One such mechanism is node activation/deactivation~\cite{KlEg02} motivated by citation networks. A set of $m$ active nodes is maintained in the model, and the new, initially active node connects to this set by $m$ links. One active node is then deactivated with probability inversely proportional to the node degree. Because of this inverse proportionality imposed by the model, the model effectively implements linear PA. Because the new connections are made to local groups of active nodes, clustering is strong. However, as shown in~\cite{VoBo03} the model is effectively one dimensional, lacking the small-world property observed in many real networks.

Another popular mechanism enforcing strong clustering is random walks~\cite{Vazquez2003,JaRo07}. A new node connects first to a random existing node, and then with some probability to one of its neighbors, and possibly to a neighbor of its neighbor, etc. Clustering is strong because the connections are concentrated in a local neighborhood of the attachment node.

\subsection{Emergent PA}

The lack of clustering is not the only problem with standard PA. Another problem is that PA {\it per se} is simply impossible in a vast majority of real networks because to ``implement'' PA, the network evolution process must ``know'' the global current structure of the whole network in order to compute the degree for each node. Since such knowledge is often unavailable in reality, PA must be an emergent phenomenon, i.e., an effective result of some other underlying evolution processes that use only local information. Yet another related problem is that such processes must lead to exactly linear PA, since if the attachment probability is not a linear function of node degree, then the degree distribution in the network is not a power law~\cite{KraReLe00}. Several mechanisms have been proposed to address these two problems as well. The aforementioned random walks, for example, do solve them both because the probabilities of the stationary distribution of a random walk on a graph are linearly proportional to node degrees. Another interesting observation was made in~\cite{FoFl06} where the authors show that connections based solely on node ranking may lead to power laws, the motivation being that node ranking is a coarser proxy to popularity than the node degree. Yet the simplest and perhaps the first model that addresses the three mentioned concerns with PA---zero-clustering, global knowledge, and linearity---is by Dorogovtsev {\it et al.}~\cite{DoMeSa01}: the new node simply selects a random existing link, and connects to its both ends. Clustering is obviously strong, and linear PA is resurrected because the probability that a random link is attached to a node of degree $k$ is proportional to $k$. However, this model is clearly a toy model, and there have been no attempts to validate it against any real networks.

\subsection{Discussion}

As far as validation is concerned, the model validation methodology is usually limited to generating synthetic graphs according to the model prescription, and comparing one or more of their structural properties, such as the degree distribution, against those in real networks. Remarkably, the core of the network evolution mechanism proposed by a model is quite rarely validated directly, because such validation is either difficult or impossible. In similarity-based models, for example, such validation is difficult because there are too many different similarity measures, and it is usually unclear which one should be used in which case~\cite{CraCo08,MaCa09}, so that cases where model predictions are validated directly against real-world similarity data~\cite{menczer02-pnas,menczer04-pnas,triangular_clustering} are rare, and usually limited to specific (types of) networks.

Within our approach, the direct validation of the network evolution mechanism is also difficult but possible. It is possible because we can infer the node coordinates in the generic similarity space as discussed in Section~\ref{sec:mapping}, and then check if the linking probability in real networks as a function of distances between nodes in this space is close to our model predictions, see Fig.~3.

In summary, the salient feature of our approach is that it {\em simultaneously}:
\begin{enumerate}
\item shows that similarity plays an important and {\em fundamental\/} role in evolution of complex networks;
\item does so by means of a very simple and {\em general\/} geometric model;
\item admits a complete {\em analytic\/} treatment;
\item {\em directly\/} validates the modeled similarity mechanism and its analytic predictions against drastically different real networks from different domains;
\item reproduces many important {\em structural\/} properties of these networks; and
\item resolves all the mentioned concerns with preferential attachment, which appears in the approach as an {\em emergent phenomenon}.
\end{enumerate}

\begin{acknowledgments}
We thank Charles Elkan, Ginestra Bianconi, Paul Krapivsky, Sid Redner, Shlomo Havlin, Eugene Stanley, and Albert-L\'aszl\'o Barab\'asi for useful discussions and suggestions. This work was supported by a Marie Curie International Reintegration Grant within the 7th European Community Framework Programme; MICINN Projects Nos.\ FIS2010-21781-C02-02 and BFU2010-21847-C02-02; Generalitat de Catalunya grant No.\ 2009SGR838; the Ram\'on y Cajal program of the Spanish Ministry of Science; ICREA Academia prize 2010, funded by the Generalitat de Catalunya; NSF Grants No.\ CNS-0964236, CNS-1039646, CNS-0722070; DHS Grant No.\ N66001-08-C-2029; and by Cisco Systems.
\end{acknowledgments}

\end{document}